\documentclass{article}
\usepackage{algorithm}
\usepackage[noend]{algpseudocode}
\usepackage{url}
\usepackage{amsmath}
\usepackage{amssymb}
\usepackage{graphicx}
\usepackage{booktabs}

\newtheorem{proposition}{Proposition}[section]

\newtheorem{proof}{Proof}
\newtheorem{problem}{Problem}

\newcommand{\ccite}[1]{\cite{#1}}

\makeatletter
\newcommand{\abs}{\@ifstar%
  \absstar%
  \absnoStar%
}
\newcommand{\norm}{\@ifstar%
  \normstar%
  \normnoStar%
}

\newcommand{\scalar}{\@ifstar%
  \scalarstar%
  \scalarnoStar%
}

\newcommand{\absstar}[2][]{%
  \def\tempa{}%
  \def\tempb{#1}%
  \ifx\tempa\tempb%
    \lvert#2\rvert%
  \else%
    \csname #1l\endcsname\lvert#2\csname #1r\endcsname\rvert%
  \fi%
}

\newcommand{\normstar}[2][]{%
  \def\tempa{}%
  \def\tempb{#1}%
  \ifx\tempa\tempb%
    \lVert#2\rVert%
  \else%
    \csname #1l\endcsname\lVert#2\csname #1r\endcsname\rVert%
  \fi%
}

\newcommand{\scalarstar}[2][]{%
  \def\tempa{}%
  \def\tempb{#1}%
  \ifx\tempa\tempb%
    \langle#2\rangle%
  \else%
    \csname #1l\endcsname\langle#2\csname #1r\endcsname\rangle%
  \fi%
}

\newcommand{\absnoStar}[1]{\left\lvert#1\right\rvert}
\newcommand{\normnoStar}[1]{\left\lVert#1\right\rVert}
\newcommand{\scalarnoStar}[1]{\left\langle#1\right\rangle}
\makeatother

\newcommand{\vect}[1]{\ensuremath{\mathbf{#1}}}

\DeclareMathOperator*{\argmax}{arg\, max}

\newcommand{\density}{\delta}   

\newcommand{\otens}{T}   
\newcommand{\ttens}{T}

\newcommand{\Mpc}{\sigma(V')}   
\newcommand{\Mpe}{\varepsilon(V')}   
\newcommand{\Mpt}{\tau(V')}

\newcommand{\avgC}[1]{\overline{w^2}(#1)}   
\newcommand{\treem}{e_b}   

\newcommand{\etal}{\textit{et al.}}

\newcommand{\latentvec}{\vect{x}}
\newcommand{\conformvec}{\vect{f}}
\newcommand{\latentmat}{\vect{X}}
\newcommand{\conformmat}{\vect{F}}

\newcommand{\probQTeam}{$T$-{\sc Comm}}
\newcommand{\probSTeam}{$T$-{\sc Team}}

\newcommand{\algCock}{{\tt Cocktail}}

\newcommand{\algQTree}[1]{{\tt CTree#1}}
\newcommand{\algQPeel}[1]{{\tt CPeel#1}}

\newcommand{\innerop}{x}
\newcommand{\expreop}{f}

\DeclareMathOperator*{\scorePeel}{score}
\DeclareMathOperator*{\scoreTree}{len}

\newcommand{\parahead}[1]{\paragraph{#1.}}

\newcommand{\DBLP}{\texttt{DBLP}}
\newcommand{\DBLPed}{\texttt{DBLP.E2}}
\newcommand{\DBLPcf}{\texttt{DBLP.C}}
\newcommand{\IMDB}{\texttt{IMDB}}

\usepackage{tikz}

\newcommand{\resultdir}{results}
\newcommand{\opsdir}{ops}

\newlength{\myheight}
\newlength{\mywidth}
\newlength{\myheightbp}
\newlength{\myheightbk}
\newlength{\mywidthbp}
\newlength{\mypwidth}
\newlength{\myheightbpV}
\newlength{\mywidthbpV}
\newlength{\mypwidthV}
\newcommand{\minmaxCT}{xmin=0.0, xmax=4,}
\newcommand{\minmaxCE}{xmin=0.0, xmax=17,}
\newcommand{\minmaxCU}{xmin=0.0, xmax=5,}

\newcommand{\minmaxTLHS}{}
\newcommand{\minmaxELHS}{}
\newcommand{\minmaxULHS}{}
\newcommand{\filebasisLHS}{}
\newcommand{\resultdirLHS}{}

\newcommand{\minmaxTRHS}{}
\newcommand{\minmaxERHS}{}
\newcommand{\minmaxURHS}{}
\newcommand{\filebasisRHS}{}
\newcommand{\resultdirRHS}{}

\newcommand{\bplotF}[5]{
}

\newcommand{\bplotR}[5]{
}

\newcommand{\bplotQ}[4]{
\begin{tikzpicture}[trim axis left, trim axis right]
\end{tikzpicture}
}

\newcommand{\bplotBlockC}[8]{
\hspace{1cm}
\begin{minipage}[t]{\mypwidth}
\bplotQ{#1}{100.0}{3}{\minmaxCU xlabel={$\Mpt$}, xticklabel pos=right, yticklabels={\algQTree{(e)}, \algQTree{(s)}, \algQPeel{(m)}, \algQPeel{(s)}, \algQPeel{(r)}}, yticklabel pos=left, xticklabels={}}
\end{minipage}
\begin{minipage}[t]{\mypwidth}
\bplotQ{#1}{100.0}{1}{\minmaxCE xlabel={$\Mpe$}, xticklabel pos=right, yticklabels={}, xticklabels={}}
\end{minipage}
\begin{minipage}[t]{\mypwidth}
\bplotQ{#1}{100.0}{0}{\minmaxCT xlabel={$\Mpc$}, xticklabel pos=right, yticklabel pos=right, yticklabels={}, xticklabels={}, ylabel={#5}}
\end{minipage}

\vspace{-1.em}

\hspace{1cm}
\begin{minipage}[t]{\mypwidth}
\bplotQ{#2}{100.0}{3}{\minmaxCU xticklabel pos=left, yticklabels={\algQTree{(e)}, \algQTree{(s)}, \algQPeel{(m)}, \algQPeel{(s)}, \algQPeel{(r)}}, yticklabel pos=left, xticklabels={}}
\end{minipage}
\begin{minipage}[t]{\mypwidth}
\bplotQ{#2}{100.0}{1}{\minmaxCE xticklabel pos=left, yticklabels={}, xticklabels={}}
\end{minipage}
\begin{minipage}[t]{\mypwidth}
\bplotQ{#2}{100.0}{0}{\minmaxCT xticklabel pos=left, yticklabel pos=right, yticklabels={}, xticklabels={}, ylabel={#6}}
\end{minipage}

\vspace{-1.6em}

\hspace{1cm}
\begin{minipage}[t]{\mypwidth}
\bplotQ{#3}{100.0}{3}{\minmaxCU xticklabel pos=left, yticklabels={\algQTree{(e)}, \algQTree{(s)}, \algQPeel{(m)}, \algQPeel{(s)}, \algQPeel{(r)}}, yticklabel pos=left, xticklabels={}}
\end{minipage}
\begin{minipage}[t]{\mypwidth}
\bplotQ{#3}{100.0}{1}{\minmaxCE xticklabel pos=left, yticklabels={}, xticklabels={}}
\end{minipage}
\begin{minipage}[t]{\mypwidth}
\bplotQ{#3}{100.0}{0}{\minmaxCT xticklabel pos=left, yticklabel pos=right, yticklabels={}, xticklabels={}, ylabel={#7}}
\end{minipage}

\vspace{-2.0em}

\hspace{1cm}
\begin{minipage}[t]{\mypwidth}
\bplotQ{#4}{100.0}{3}{\minmaxCU xticklabel pos=left, yticklabels={\algQTree{(e)}, \algQTree{(s)}, \algQPeel{(m)}, \algQPeel{(s)}, \algQPeel{(r)}}, yticklabel pos=left}
\end{minipage}
\begin{minipage}[t]{\mypwidth}
\bplotQ{#4}{100.0}{1}{\minmaxCE xticklabel pos=left, yticklabels={}}
\end{minipage}
\begin{minipage}[t]{\mypwidth}
\bplotQ{#4}{100.0}{0}{\minmaxCT xticklabel pos=left, yticklabel pos=right, yticklabels={}, ylabel={#8}}
\end{minipage}

\vspace{-1.5em}
}

\newcommand{\hplotOpsX}[2]{
\begin{tikzpicture}[trim axis left]
\end{tikzpicture}
}

\newcommand{\hplotOpsN}[3]{
\begin{tikzpicture}[trim axis left]
\end{tikzpicture}
}

\newcommand{\hplotOps}[2]{
\begin{tikzpicture}[trim axis left]
\end{tikzpicture}
}

\usetikzlibrary{external}
\begin{document}

\title{Finding low-tension communities}

\author{
Esther Galbrun \\ Inria Nancy -- Grand Est, France \\ \url{esther.galbrun@inria.fr}
\and 
Behzad Golshan \\ Recruit Institute of Technology, CA, USA \\ \url{behzad@recruit.ai}
\and 
Aristides Gionis \\ Aalto University, Finland \\ \url{aristides.gionis@aalto.fi}
\and
Evimaria Terzi \\ Boston University, MA, USA \\ \url{evimaria@cs.bu.edu}
}

\date{}

\maketitle

\begin{abstract}
Motivated by applications that arise in online social media and
collaboration networks, there has been a lot of work on
\emph{community-search} and \emph{team-formation} problems.  In the
former class of problems, the goal is to find a subgraph that
satisfies a certain connectivity requirement and contains a given
collection of seed nodes. In the latter class of problems, on the
other hand, the goal is to find individuals who collectively have the
skills required for a task and form a connected subgraph with certain
properties.

In this paper, we extend both the community-search and the
team-formation problems by associating each individual with a
\emph{profile}. The profile is a numeric score that quantifies the
position of an individual with respect to a topic. We adopt a model
where each individual starts with a \emph{latent profile} and arrives
to a \emph{conformed profile} through a dynamic \emph{conformation
  process}, which takes into account the individual's social
interaction and the tendency to conform with one's social environment.
In this framework, \emph{social tension} arises from the differences
between the conformed profiles of neighboring individuals as well as
from differences between individuals' conformed and latent profiles.

Given a network of individuals, their latent profiles and this
conformation process, we extend the community-search and the
team-formation problems by
requiring the output subgraphs to have low social tension.  From the
technical point of view, we study the complexity of these problems and
propose algorithms for solving them effectively.  Our experimental
evaluation in a number of social networks reveals the efficacy and
efficiency of our methods.


\end{abstract}

\section{Introduction}
%
%
 
A large body of work in social and collaboration networks focuses on
solving variants of two classes of problems: \emph{community-search}
and \emph{team-formation}.  The high-level goal of both classes is to discover a connected subgraph, but they differ in the input query.
In the community-search
problem~\ccite{faloutsos04fast,DBLP:conf/sigmod/RuchanskyBGGK15,DBLP:conf/kdd/SozioG10,tong06center},
the input query consists of individuals who should participate in the
subgraph.  
In the team-formation
problem~\ccite{DBLP:conf/cikm/AnagnostopoulosBCGL10,Anagnostopoulos:2012:OTF:2187836.2187950,feng14search,lappas09finding,DBLP:conf/sdm/BhowmikBGP14,DBLP:conf/sdm/GajewarS12,Kargar:2011:DTT:2063576.2063718,5590958,Rangapuram:2013:TRT:2488388.2488482}, on the other hand, the input query consists of a set of skills that need to be covered by the solution.

These two problem classes have applications primarily in
online social media and collaboration networks.
For example, solutions to the community-search problem can be used to identify a set of individuals 
who are the most appropriate group to organize a social gathering.  
Similarly, solutions to the team-formation problem can be used
by project managers or human-resource (HR) departments of companies and universities in order 
to identify a well-functioning team of employees for a project
or to make a competitive cluster hire. 
Such hiring and resource-allocation decisions can prove important for the well-functioning and the success of an organization.

In all existing variants of the community-search and the
team-formation problems the goal is to discover a connected subgraph;
Different variants of the problem impose different
requirements on the structure of the solution subgraph.  For example, in some
settings one asks to find high-density
communities~\ccite{DBLP:conf/kdd/SozioG10}, while in others the objective is
to find small-diameter communities~\ccite{DBLP:conf/sigmod/RuchanskyBGGK15}.
From the computational point of view, different requirements lead to 
different problems that have different complexities and require the design of different algorithms.
As usual, the choice of the appropriate problem formulation depends on the application domain.

In contrast to the static requirements imposed by existing work, 
in this paper we incorporate the dynamics of social interactions in the community-search problem. 
We do so by associating \emph{profiles} to the individuals in the network.

Regarding any subject of matter, individuals have their own opinion or
preference. However, in a social context, such as a working group or
team, the opinion that is publicly held or enacted by each individual is
influenced by his peers. Diverging from the opinion of peers generates
a social cost bearing on the individual. Simultaneously, expressing an
opinion that does not match one's own beliefs creates an inner
conflict and an associated internal cost.  These constitute important
factors impacting the operation of the community as a social group.
Therefore, in this paper, we extend the community-search and the
team-formation problems to take them into account, adding a new and
realistic perspective.

We consider a social network between individuals, each one of which is
associated with a \emph{profile}.  
The profiles of individuals model
their interests, skills, preferences, opinions, and so on, with
respect to different aspects or topics.  For example, 
the profile may represent the interest of an individual in discussing
about politics, or the working style of an individual --- e.g.\ whether he is a morning person or a night owl.
Since profiles may cover a number of
different aspects or topics, we assume that they are represented as
multi-dimensional vectors.

We further assume that 
the profiles of individuals change due to the social influence they receive from other individuals in the network.
We model this change through what we call \emph{conformation process}, which is a dynamic 
process.  
Motivated by existing work on models of opinion formation and social
influence~\ccite{Clifford1973,DeGroot74,doi:10.1080/0022250X.1990.9990069,Holley1975,friedkin99social} 
we assume that the conformation process is a repeated-averaging process.
In our work, we generalize this process from one-dimensional opinions to multi-dimensional profiles.
The effect of this process is that the initial profiles of individuals, which we call \emph{latent profiles},
get re-enforced or altered through synthesis and aggregation of different viewpoints of the  network participants.

Given this process, the latent profiles of individuals and a social network that represents their social interaction, \emph{social tension} arises because of the
differences between the conformed profiles of neighboring nodes and
between the conformed and latent profiles of the nodes themselves. In this context, our
goal is to identify communities and teams that are not only
connected and qualified, but also exhibit low social tension.
We refer to these problems as the 
 {\probQTeam} and the {\probSTeam} problems respectively.  
To the best of our knowledge we are the first to define and study these problems in the light of 
profile-conformation processes in a social network.

In terms of technical results, we show that both the {\probQTeam} and the {\probSTeam} problems
are NP-hard and we design graph-theoretic algorithms for solving them. A key difficulty that we overcome while
designing these algorithms is that we do not run the conformation process 
for every step of these algorithms; indeed, this would be
computationally very demanding. Rather, we create effective 
proxies of this process that allow our algorithm to scale.
Our experiments with real-world data demonstrate the efficiency and the efficacy of our 
algorithms for both problems.

\parahead{Applications} 
Both the {\probQTeam} and {\probSTeam} problems 
have numerous potential applications.  
For example, when analyzing social networks or social media it is often
useful to identify connected groups of users who have similar profiles with respect to a topic or an idea 
(i.e.\ they are in agreement). Note, however, that minimizing
the social tension does not necessarily imply looking for highly
homogeneous communities but, rather, favoring communities that are able to bridge
opinion gaps at low social and communication costs.  Groups with low
social tension can be recommended as ideal in order to form a 
group to organize or to invite at a social event.  Such problems also
arise in human-resource management, when searching for groups of
workers who can collaborate together in a conflict-free manner in
order to successfully complete a project.  Being able to identify a
group of people who are not going to experience high social tension is
particularly useful when considering cluster hires in universities or
start-up companies, where the investment is high and the human factor risk needs to be minimized.

\parahead{Roadmap}  
The rest of the paper is organized as follows.
In Section~\ref{sec:related} we give a thorough
overview of the related work and identify the connections to our own work. 
After describing our notation and modeling
assumptions in Section~\ref{sec:preliminaries}, 
we proceed to  Section~\ref{sec:qproblem}, 
where we define the community-search variant of our low-tension mining problem, 
prove that it is NP-hard and present algorithms for solving it in practice.
We discuss other problem variants in Section~\ref{sec:variants}.
Our experimental results are described in Section~\ref{sec:exp} 
and we conclude the paper in Section~\ref{sec:conclusions}.

A short version of this paper appeared in SDM'17. In this extended
version, we discuss the team-formation problem variant, beside the
original community-search problem, and include additional experimental
results.

\section{Related Work}
\label{sec:related}
To the best of our knowledge, we are the first to combine the problem
of identifying a group of nodes from an input graph with an underlying
dynamic conformation process of opinion formation at play over the graph.
However, while the {\probQTeam} and {\probSTeam} problems as we define them are novel,
our work is related to existing work on graph mining and opinion
dynamics. We summarize the related literature below.

\subsection{Community search} 
Given a graph and a subset of its nodes as
a seed, \emph{community-search} problems ask for a set of nodes which 
is a superset of the seed nodes and
induces a connected subgraph in the original graph. 
Since this class of problems was initially introduced~\ccite{faloutsos04fast,tong06center},
different instances of the community-search problem appeared. Each one of them imposes a different requirement on 
the graph-theoretic properties of the reported subgraph. 
For example, Sozio and Gionis~\ccite{DBLP:conf/kdd/SozioG10} focus on
finding a subgraph that connects the chosen seed members and has
maximum density, Ruchansky
\etal~\ccite{DBLP:conf/sigmod/RuchanskyBGGK15} solve the same problem
with the objective to minimize the sum of pairwise shortest path
distances among the nodes participating in the subgraph. Faloutsos
\etal~\ccite{faloutsos04fast} study the problem of finding a subgraph
that links two seed nodes while maximizing the quality of connection
of the subgraph, for a generic measure of quality.  Tong
\etal~\ccite{tong06center} later extended their work to an arbitrary
number of seed nodes.

Although these problems are related to the {\probQTeam} problem we study here, 
the novelty of our problem comes from the fact that we associate nodes with profiles and that 
we assume  
a profile-conformation process that takes place over the network. Our objective function is directly related
to this process as it aims to minimize the social tension in the reported community.
As a result, the technical results 
we obtain for {\probQTeam} are different from the other community-search problems.


\subsection{Team formation}  
\emph{Team-formation} is another class of graph-mining problems where
the goal is to find a set of individuals in a network,
who together possess the required skills while minimizing the communication
costs between the members of the team. Since the introduction of this problem
in the domain of data mining by Lappas \etal~\ccite{lappas09finding}, a number
of variants have been studied, 
from varying the communication cost
function~\ccite{DBLP:conf/sdm/BhowmikBGP14,DBLP:conf/sdm/GajewarS12,Kargar:2011:DTT:2063576.2063718,5590958,Rangapuram:2013:TRT:2488388.2488482}, to considering the workload
distribution~\ccite{DBLP:conf/cikm/AnagnostopoulosBCGL10}, designing
algorithms for online team
formation~\ccite{Anagnostopoulos:2012:OTF:2187836.2187950}, and maximizing
the influence of the team over the network~\ccite{feng14search}.

At a high level, the {\probSTeam} problem we define is a team-formation problem.
However, all existing work on team-formation problems aims to identify teams that optimize some static measure (e.g.\
the density or the diameter) of the subgraph that the team induces.  However, the existence of 
a dynamic conformation process over the network makes our problem definition and setting quite distinct.
Our goal is to minimize the social tension, which is much more complicated as it is caused
by
the difference in the profiles which arise from the interplay between the conformation process and the network structure.
Therefore, existing algorithms for team formation cannot be applied to our problem.




\subsection{Opinion dynamics}  
Starting in the 1970s, models have been built that try to capture
the opinion-formation processes in groups and networks~\ccite{Clifford1973,DeGroot74,doi:10.1080/0022250X.1990.9990069,Holley1975,Yildiz:2013:BOD:2542174.2538508,friedkin99social}. For example, 
\emph{voter models}, pioneered by 
Clifford and Sudbury~\ccite{Clifford1973}
and 
Holley and Liggett~\ccite{Holley1975}
 are stochastic models of opinion formation where at each step a node is selected at random and adopts the opinion of one of its neighbors, also selected at random.
In DeGroot's \emph{averaging model}~\ccite{DeGroot74}, each node updates its opinion, 
by the weighted average of its own opinion and its neighbors' opinion at the previous time step.
Friedkin and Johnsen~\ccite{doi:10.1080/0022250X.1990.9990069,friedkin99social} introduced a model where every node has an immutable inner opinion and a changeable expressed opinion.
Each node forms its expressed opinion in a repeated-averaging process involving its own inner opinion and the expressed opinions of its neighbors. 
Given the popularity of this model we also adopt
it for modeling our profile-conformation process, and extend it to multiple dimensions.

The Friedkin and Johnsen model has been used in recent works by 
Bindel {\etal}~\ccite{Bindel:2011:BFY:2082752.2082924} and by 
Gionis {\etal}~\ccite{DBLP:conf/sdm/GionisTT13}. Bindel {\etal}\ focus
on the price of anarchy in terms of the tension achieved through local repeated averaging
and global opinion coordination. Gionis {\etal}\ aim at 
identifying the set of nodes whose opinions need to be changed so that the overall positive
opinion in the network is maximized. Although our work builds upon Friedkin
and Johnsen's ideas to model the profiles and their conformation process, none
of the above-mentioned works addresses the question of identifying a subgraph with low
social tension as we do.


Extending our work to other times of opinion-dynamics models, i.e.\ beyond the model of Friedkin and Johnsen, is a promising direction for future research.

\section{Preliminaries}
\label{sec:preliminaries}

Throughout the paper we consider a social network $G=(V,E)$ where the nodes
in $V$ correspond to individuals and the edges in $E$ represent the interactions
between these individuals. For simplicity of exposition, 
we present the case of an unweighted and undirected graph, 
but the problems and algorithms we discuss can naturally be extended 
to weighted and directed graphs. 

The set of neighbors of node $i$ in $G$ is denoted by $N_G(i)$. Given a subset of
vertices $U \subseteq V$, we let $E(U)$ denote the set of edges of $G$ \emph{induced}
by $U$, i.e.\ $E(U) = \{(i,j) \in E, i,j \in U \}$, and $G(U)$ denote the
corresponding \emph{induced subgraph} $G(U) = (U,E(U))$.

\subsection{Profiles} 
Evidently, each individual has their own set of preferences (e.g.\  style, habits, biases,
and opinions). We refer to this personal set of preferences as a \emph{profile}.
For now, let us assume that profiles only reflect preferences regarding a single
aspect (e.g.\ working style), that is, profiles consist of a single attribute. 
We assume that profile attribute values are represented by a real number in 
the interval $[0, 1]$.
For instance, a value between $0$ and $1$ may represent an individual's
preference towards a certain software tool, or
his preference to work in a team or in isolation respectively.

The key characteristic of our model is that it captures the
interaction between the user profiles and their social
connections. This is done by assuming that each individual $i$ has a
\emph{latent profile} and a \emph{conformed profile}, denoted by
$\innerop_i$ and $\expreop_i$ respectively. The latent profile of an
individual represents the individual's own true preference. However,
individuals may choose \emph{not} to act in accordance with their
latent profiles as they try to minimize peer pressure by conforming
their preferences to those of their peers. The conformed profile
represents these adjusted preferences.

For simplicity, we first describe the model for single-attribute
profiles, but later on we discuss how to extend it to multi-attribute
profiles.  We summarize the latent and conformed profiles of all $n$
individuals with respect to a single attribute using vectors
$\latentvec$ and $\conformvec$ respectively.

\subsection{Measuring tension}
Due to the underlying social structures and mechanisms, the
conformed profile $\expreop_i$ of a node can differ from its latent profile
$\innerop_i$. 
In such case,  the node will bear an \emph{inner tension}
caused by the difference between its own latent and conformed profiles.
On the other hand, the difference between the node's conformed profile and between the conformed profiles of its neighbors will
cause \emph{cross tension}.  Hence, the total tension on node $i$ is:
\begin{equation*}
\label{eq:mini}
\otens_i(G, \latentvec, \conformvec) = (\innerop_i-\expreop_i)^2 + \sum_{j \in N_G(i)} (\expreop_i-\expreop_j)^2.
\end{equation*}
Then, the \emph{social tension} of the network
is simply the sum of the individual tensions, defined as
\begin{align}
\label{eq:overall-tension}
\ttens(G, \latentvec, \conformvec) & =  \sum_{i \in V} \otens_i(G, \latentvec, \conformvec) \nonumber \\
                                   & =  \sum_{i \in V} \big( (\innerop_i-\expreop_i)^2 + \sum_{j \in N_G(i)} (\expreop_i-\expreop_j)^2 \big),
\end{align}
which can alternatively be written as the sum of the overall inner and cross tensions
\begin{equation}
\ttens(G, \latentvec, \conformvec)
= \sum_{i \in V} (\innerop_i-\expreop_i)^2 + \sum_{(i,j) \in E} 2 \, (\expreop_i-\expreop_j)^2. \nonumber
\end{equation}

\subsection{Conformation process}
But how do nodes arrive at their conformed profiles? Consider a repeated averaging
process where at each step each node adjusts its conformed profile by setting
it to the average of its latent profile and the conformed profile of its neighbors.
Formally, denoting as $\expreop_i(t)$ the conformed profile of node $i$ at
step $t$, we have:
\begin{equation}\label{eq:repeatedaveraging}
    \expreop_i(t+1) = \frac{\innerop_i + \sum_{j \in N_G(i)} \expreop_j(t)}{1 + \abs{N_G(i)}}.
\end{equation}

Computing the conformed profiles according to the \emph{repeated averaging model}
is equivalent to choosing $\expreop_i$ to minimize $\otens_i(G, \latentvec, \conformvec)$.
That is, if each node aims to minimize its tension, the \emph{repeated averaging
model} provides an optimal choice for the conformed profiles. In that sense,
using the \emph{repeated averaging model} yields a Nash equilibrium for the
tension, not a social optimum~\ccite{Bindel:2011:BFY:2082752.2082924}.

A practical consideration is that in this model the latent profiles are assumed to be known, 
while the conformed profiles are the output of the conformation process. 
But, in practice, we have access to the conformed profiles while the latent profiles cannot be observed. 
This, however, does not constitute a problem for our model. 
One can swap the known and unknown variables of the system
and solve for the latent profiles, given the conformed profiles and the original network.

The model adopted here for how conformed profiles emerge is a well-studied
opinion-formation and social-influence model that has been introduced by sociologists
~\ccite{DeGroot74,doi:10.1080/0022250X.1990.9990069,friedkin99social}. In particular,
the work of Friedkin {\etal}~\ccite{friedkin99social} validates this model by conducting
a set of controlled experiments in which they observe how interactions between
individuals in small groups influence their expressed opinions. The study
demonstrates that the repeated-averaging model can both predict the opinions
that individuals converge to, as well as the rate of convergence. Others
have studied the mathematical properties of this model.
For instance, it has been shown~\ccite{Bindel:2011:BFY:2082752.2082924,DBLP:conf/sdm/GionisTT13}
that the process 
converges in polynomial time to a fixed-point
solution. In fact, the final conformed profiles can be computed by a matrix
inversion~\ccite{DBLP:conf/sdm/GionisTT13}, but actually repeating the averaging
process leads to much faster computation. 

\subsection{Multi-attribute profiles}
Our assumption so far has been that profiles reflect the preferences of individuals
with respect to a single attribute. But our notion of latent and conformed profiles can
be easily extended to the case of multiple attributes, leading to multi-attributes 
--- i.e.\ multi-dimensional ---
profiles. Assume there are $m$ aspects or topics of interest, each associated to an attribute. 
The latent and conformed
profiles can be simply extended from a single real number to real-valued vectors
of dimensionality $m$, where each entry corresponds to one of the attributes. 
We summarize the latent and conformed profiles of $n$ individuals on $m$ attributes using
$n \times m$ matrices $\latentmat$ and $\conformmat$ respectively. Note that
each column of these matrices, denoted by $\latentvec_a$ and $\conformvec_a$ for $a=1,\ldots, m$,
corresponds to a single attribute.  In the multi-attribute case, the conformed
profiles can be computed as before by applying Equation~\eqref{eq:repeatedaveraging}
in a column-wise fashion. Similarly, the social tension $\ttens(G, \latentmat, \conformmat)$ is defined as the sum of the social tensions across all $m$ attributes.


\section{Finding low-tension communities with seed nodes}
\label{sec:qproblem}

In this section, we introduce the {\probQTeam} problem, study its complexity and provide algorithms for solving it.

\subsection{Definition of the {\large {\probQTeam}} problem}
\label{ssec:qproblem}
At a high level, the {\probQTeam} problem aims to find a connected, low-tension
community that involves a chosen subset of members. Formally, this intuitive statement
is captured by the following problem definition:
\begin{problem}[{\probQTeam}]
Given a network $G = (V,E)$, latent profiles $\latentmat$ and a set of seed nodes
$Q\subseteq V$, find $V'\subseteq V$ such that $Q\subseteq V'$, the graph $G'$
induced by $V'$ on $G$ is connected and $\ttens(G', \latentmat, \conformmat)$ is
minimized, where $\conformmat$ is computed by the repeated averaging model on
$G'$.
\end{problem}

Note that when defining the {\probQTeam} problem, we assume that the social
tension is computed as in Equation~\eqref{eq:overall-tension}, $\latentmat$ is
the matrix containing the latent profiles of the nodes in $V'$ and $\conformmat$
contains the conformed profiles of individuals, computed using the repeated
averaging model described in the previous section (see Equation~\eqref{eq:repeatedaveraging})
over the subgraph induced by $V'$.

Given these assumptions, we can make the following observations with respect to
the requirements imposed on the solution of {\probQTeam}: as the number of edges
in the resulting subgraph and the differences in the conformed profiles across
these edges decrease, so does the social tension. In particular, the complete
absence of edges results in no tension at all. However, the requirement that the
output subgraph should be connected forbids such solutions. From the application
point of view, connectivity is important as it guarantees communication among
the community members. One can see that minimizing tension and guaranteeing connectivity
leads to an interesting trade-off between the density of edges and the homogeneity of the
profiles of nodes in the reported subgraph. Communities should consist of individuals
who share similar profiles or individuals who have
divergent profiles but are needed to guarantee connectivity, these latter ones
being preferably very sparsely connected with the rest of the community members.

with respect to the computational complexity of the {\probQTeam} problem, we obtain the following result.

\begin{proposition}\label{prop:hardness}
The {\probQTeam} problem cannot be solved optimally in polynomial time even with a single attribute  
{\em (}i.e.\ $m=1${\em )} unless $\text{P}=\text{NP}$.
\end{proposition}


\begin{proof} (Sketch)
We prove the hardness of  \probQTeam{}  with a reduction from the
problem of exact cover by 3-sets.

The exact cover by 3-sets problem, \textbf{X3C} for short, asks the following
question. Given a universe of elements $B = \{ b_1, \dots, b_p \}$,
where the number of elements $p$ is a multiple of $3$, and a collection
$S = \{s_1, \dots, s_q \}$ of 3-elements subsets of $B$, is there a
collection $S' \subseteq S$ such that every element in $B$ occurs in exactly
one member of $S'$?

Given an instance of the \textbf{X3C} problem, we construct an instance of the
\probQTeam{} problem with one-attribute profiles (i.e.\ m = 1) as follows. To
each element $b_i$ in $B$ we associate a node, called an ``element-node'', with
latent profile $0$, and to each set $s_j$ in $\mathcal{S}$ we associate a node,
called a ``set-node'', with latent profile $1$. Each set-node is connected to the
three nodes that represent the elements it contains. In addition, we make all the
$q$ set-nodes part of a larger clique $D$ of $o+q$ nodes with latent profiles
equal to $1$. Also, we make each element-node $b_i$ part of a larger clique
$A_i$ containing $o$ nodes with latent profiles $0$. Finally, we assume that all
the nodes in our construction are seed nodes except for the $q$ set-nodes.
This construction is illustrated in Figure~\ref{fig:redu}.

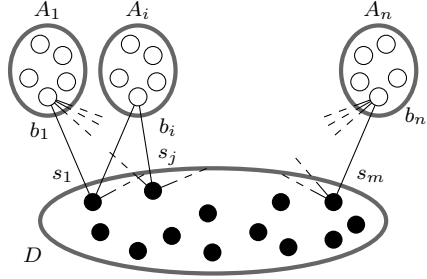
\begin{figure}[t]
\centering
\begin{tikzpicture}
\footnotesize
\draw[black!60, ultra thick] (-2.2,2) ellipse (0.5cm and 0.6cm);
\node at (-2.2,2.8) {$A_1$};
\node at (-2.3,1.2) {$b_1$};
\node at (-2,.6) {$s_1$};
\draw[black!60, ultra thick] (-1,2) ellipse (0.5cm and 0.6cm);
\node at (-1,2.8) {$A_i$};
\node at (-0.6,1.25) {$b_i$};
\node at (-.6,.85) {$s_j$};
\draw[black!60, ultra thick] (2.2,2) ellipse (0.5cm and 0.6cm);
\node at (2.2,2.8) {$A_n$};
\node at (2.7,1.4) {$b_n$};
\node at (2.1,.6) {$s_m$};
\draw[black!60, ultra thick] (0,0) ellipse (2.3cm and 0.7cm);
\node at (-2.4,-.45) {$D$};
\draw (-1.6,.25) -- (-2.2,1.65);
\draw (-1.6,.25) -- (-1,1.65);
\draw (-.8,.4) -- (-1,1.65);
\draw (1.6,.25) -- (2.2,1.65);

\draw[ddotfade] (1.6,.25) -- (.4,1.65);
\draw[ddotfade] (1.6,.25) -- (-1.,1.65);
\draw[ddotfade] (-1.6,.25) -- (1.,1.65);
\draw[ddotfade] (-.8,.4) -- (2.2,1.65);
\draw[ddotfade] (-.8,.4) -- (-2.2,1.65);

\draw[ddotfade] (-2.2,1.65) -- (1.6,.25);
\draw[ddotfade] (-2.2,1.65) -- (-.8,.4);
\draw[ddotfade] (-2.2,1.65) -- (.3,.3);

\draw[ddotfade] (2.2,1.65) -- (-0,.4);
\draw[ddotfade] (2.2,1.65) -- (-1.6,.25);
\draw[ddotfade] (2.2,1.65) -- (.6,.25);

\fill[black] (-1.6,.25) circle (.12cm);
\fill[black] (-.8,.4) circle (.12cm);
\fill[black] (1.6,.25) circle (.12cm);
\fill[black] (-1.5,-.15) circle (.12cm);
\fill[black] (-1.,-.4) circle (.12cm);
\fill[black] (-.55,-.2) circle (.12cm);
\fill[black] (-.07,.1) circle (.12cm);
\fill[black] (-.01,-.42) circle (.12cm);
\fill[black] (1,-.35) circle (.12cm);
\fill[black] (1.9,-0.05) circle (.12cm);
\fill[black] (1.6,-0.25) circle (.12cm);
\fill[black] (0.6,-0.15) circle (.12cm);
\fill[black] (0.9,0.25) circle (.12cm);

\filldraw[fill=white, draw=black] (-2.2,1.65) circle (.12cm);
\filldraw[fill=white, draw=black] (-1,1.65) circle (.12cm);
\filldraw[fill=white, draw=black] (2.2,1.65) circle (.12cm);

\filldraw[fill=white, draw=black] (-2.3,2.35) circle (.12cm);
\filldraw[fill=white, draw=black] (-2.45,1.9) circle (.12cm);
\filldraw[fill=white, draw=black] (-2.,2.2) circle (.12cm);
\filldraw[fill=white, draw=black] (-1.95,1.85) circle (.12cm);

\filldraw[fill=white, draw=black] (-1.1,2.35) circle (.12cm);
\filldraw[fill=white, draw=black] (-1.25,1.9) circle (.12cm);
\filldraw[fill=white, draw=black] (-.8,2.2) circle (.12cm);
\filldraw[fill=white, draw=black] (-.75,1.85) circle (.12cm);

\filldraw[fill=white, draw=black] (2.3,2.35) circle (.12cm);
\filldraw[fill=white, draw=black] (2.45,1.9) circle (.12cm);
\filldraw[fill=white, draw=black] (2.,2.2) circle (.12cm);
\filldraw[fill=white, draw=black] (1.95,1.85) circle (.12cm);
\end{tikzpicture}
\caption{Schema of the constructed graph for the reduction from \textbf{X3C}. The
black and white nodes have latent profiles with values 1 and 0 respectively.}
\label{fig:redu}
\end{figure}

We prove that if the given instance of \textbf{X3C} has an exact 3-set cover,
then the minimum tension subgraph solution of \probQTeam{} will contain 
the set-nodes of this exact cover.

The main idea of the proof is based on the following observation. Selecting
a large (yet polynomial in $p$) value for $o$, increases the size of
all cliques in our construction (i.e.\ clique $D$ and all $A_i$ cliques).
As a result, the conformed profile of all the nodes (including the set-nodes)
in $D$ will be very close to $1$, and the conformed profiles of all nodes in
the $A_i$ cliques will be very close to $0$. Thus, the only non-negligible source
of tension will be the tension across the edges that connect element-nodes
to set-nodes. Each such edge would increase the tension by almost a unit.

Note that since each $b_i$ is a seed node, it has to be connected to the $o$
seed nodes in $D$. This can be achieved only by going through a set-node. Thus,
the solution to \textbf{X3C} has to pick a subset of set-nodes that cover
all the element-nodes to ensure connectivity. Now, if $x$ set-nodes are
included in the subgraph then the tension would be roughly $3x$. Obviously, the
solution that minimizes the tension is the solution that picks the smallest number
of set-nodes. This is achieved by selecting an exact cover, if such cover exists.
\end{proof}

\subsection{Algorithms for the {\large {\probQTeam}} problem}
\label{ssec:algorithms}


While the objective of the {\probQTeam} problem is to minimize
the social tension in the solution graph, $\ttens(G', \latentmat, \conformmat)$,
obtaining the conformed profiles through the repeated-averaging process is costly;
thus, it is not feasible to compute the social tension on a large number of
candidate subgraphs. A possible alternative is to compute the conformed
profiles by applying the repeated-averaging process on the whole graph once and use
these profiles when evaluating the tension in the candidates later on.
However, while designing our algorithms, we observed that this is a poor
choice. Intuitively, the presence of a node with a latent profile that departs
greatly from its neighbors' might significantly sway their conformed profiles.
During the search, this node will likely be removed from the candidate set early
on, but  its effect would remain.

In order to avoid such effects, but also 
avoid the repeated computations of social tension we use the following trick in all 
our algorithms: 
for 
a pair of neighboring nodes $i$ and $j$ we assign to edge $(i,j)\in E$ weight
$w_{ij} = \left |\innerop_i - \innerop_j\right |$. We then use this weight as a way of quantifying the contribution
of this edge in the overall social tension.  Although $w_{ij}$ is just a proxy of the edge's contribution, the trick appears
to perform well in practice and leads to significant speedups.
Once our algorithms obtain the set of nodes to report, we apply the
repeated-averaging process on the induced subgraph in order to evaluate the
social tension of the solution.

We propose two approaches for finding good candidate solutions for this problem.

\parahead{Spanning-tree approach}
This approach connects the seed nodes by building a spanning tree between
them and is based on the $2$-approximation algorithm for 
the Steiner tree problem~\ccite{vazirani03approximation}. 
It works as follows: first, it computes
the shortest path between every pair of seed nodes. Next, it constructs a complete
graph over the seed nodes such that the weight of the edge between two nodes
corresponds to their shortest path distance in the original graph. Then, it
considers the minimum-spanning tree from this complete graph, e.g.\ obtained with
Prim's algorithm, and replaces each edge of the spanning tree with the associated
original shortest path. The resulting subgraph constitutes the output of our
tree-based algorithm. A sketch of this algorithm, which we call \algQTree{}, is shown
in Figure~\ref{alg:QT}.
Note that this approach, searching for the best spanning tree, lacks any control over the induced edges that will be included in the solution. In this sense, it is an optimistic strategy.

We obtain different variants of this algorithm depending on the measure used to
evaluate the length of a path. Having experimented with various options, we focus
on two variants, where the length of the path is either 
\vspace{-.3em}
\begin{itemize}
\setlength{\topsep}{0pt}
\setlength{\itemsep}{-.2em}
\setlength{\parsep}{0pt}
\item [($i$)] the number of edges involved, or
\item [($ii$)] the sum of weights of the edges along the path. 
\end{itemize}
In other words, if $p_{ij} = (i, v_1, v_2, \dots, v_k, j)$ is a path between $i$ and $j$, 
we have
$\scoreTree(p_{ij}) = k + 1$ in the first variant, and
$\scoreTree(p_{ij}) = w_{iv_1} + w_{v_1v_2} + \dots + w_{v_kj}$ in the other. 
We denote these variants as \algQTree{(e)} and \algQTree{(s)} respectively.

\begin{figure}
\centering
\begin{minipage}[c]{.66\textwidth}
\input{pseudo_code_QT}
\end{minipage}
\caption{The \algQTree{} algorithm for solving \probQTeam{}.}
\label{alg:QT}
\end{figure}

The main step in our algorithm is the computation of the shortest path between
all pairs of seed nodes by running the Dijkstra algorithm from each seed node in
turn. The running time of our algorithm is thus ${\mathcal{O}}(\abs{Q}(\abs{V} + \abs{E} \log \abs{E}))$.

\parahead{Top-down approach}
The second algorithm is a top-down approach which starts with the full graph and
iteratively removes nodes until it is no longer possible to continue without
disconnecting the seed nodes. The pseudo-code for this algorithm, called
\algQPeel{}, is given in Figure~\ref{alg:QP}.

Again, we obtain different variants, this time by varying the $\scorePeel()$ function
 for choosing the next node to remove. We selected the following three
scores:
\vspace{-.3em}
\begin{itemize}
\setlength{\topsep}{0pt}
\setlength{\itemsep}{-.2em}
\setlength{\parsep}{0pt}
\item [($i$)] The score is a number assigned randomly to each node when initializing the
algorithm, which determines the order in which the nodes are peeled. This random
variant is denoted as \algQPeel{(r)}.
\item [($ii$)] The score is the sum of the weights of the remaining edges adjacent to the node,
i.e.\
\setlength{\itemsep}{-.8em}
\vspace{-.5em} $$\scorePeel(i, U) = \sum_{j \in N_{G(U)}(i)} w_{ij}.$$
\item [($iii$)] The score is the largest weight among the remaining edges adjacent to the node,
i.e.\
\vspace{-.5em} $$\scorePeel(i, U) = \max_{j \in N_{G(U)}(i)} w_{ij}.$$
\end{itemize}

The second and third scores are similar but the former uses a sum where
the latter takes the maximum, resulting in variants \algQPeel{(s)} and
\algQPeel{(m)}, respectively. Nodes with larger scores are considered
first (line~\ref{alg:QP_pick} in Figure~\ref{alg:QP}). Meanwhile, in
all three variants, nodes that get isolated are pruned away
(line~\ref{alg:QP_trim}).  Rather than favoring good connections, this
second approach removes nodes that might generate large costs, and
thus follows a pessimistic strategy.

In practice, we can find a minimum connecting tree using the strategy described
previously. Then, when we are about to remove a node, we check whether it belongs
to the current tree, in which case we need to look for an alternative tree that
does not involve this node. Only if such a tree can be found can we safely remove
the node. In the worst case, we would have to recompute the spanning tree for
each node, resulting in a running time ${\mathcal{O}}(\abs{V}(\abs{V} + \abs{E} \log \abs{E}))$.

\begin{figure}
\centering
\begin{minipage}[c]{.66\textwidth}
\input{pseudo_code_QP}
\end{minipage}
\caption{The \algQPeel{} algorithm for solving \probQTeam{}.}
\label{alg:QP}
\end{figure}

\section{Problem Variants}\label{sec:variants}
In this part, we discuss some variants of the {\probQTeam} problem that make
different assumptions with respect to the input, while the output is
always a connected subgraph with low social tension.

\subsection{Finding low-tension teams with chosen skills} 
In many team-formation
problems, there is a universe $S$ of \emph{skills} that individuals may possess
and each individual $i$ is associated with a set of skills $S_i\subseteq S$. 
Note that the skills are different from the attributes of the profiles; skills are Boolean features associated with
individuals (i.e.\ an individual has the skill or not) 
and do not change because of social interactions.  In contrast, profiles are real-valued and the conformed profiles are subject to change due to social pressure.
	
\parahead{Definition of the {\probSTeam} problem}
Given
a project requiring a subset $P \subseteq S$ of skills,
the problem is to find a subset of individuals $V'\subseteq V$ such that for every
skill required by the project there is at least one individual in $V'$ who
possesses it. When individuals are organized in a network, it is also required that the subgraph induced
by $V'$ satisfies certain properties that capture the ease of communication
within the team, e.g.\ small diameter, low weight spanning tree, 
small density, etc.~\cite{DBLP:conf/sdm/GajewarS12,Anagnostopoulos:2012:OTF:2187836.2187950,lappas09finding}.

In our setting, we consider a version of the team-formation problem where the
goal for $V'$ is to be such that the required skills are covered, the graph $G'$
induced by $V'$ is connected and the social tension in $G'$ is minimized.
This problem can be formally defined as follows.
 
\begin{problem}[{\probSTeam}]
Given a network $G = (V,E)$ and latent profiles $\latentmat$, as well as a universe
of skills $S$, a set of skills $S_i\subseteq S$ for every $i \in V$ and a project
requiring a subset of skills $P \subseteq S$, find $V'\subseteq V$ such that
$P \subseteq \bigcup_{i\in V'} S_i $,
the graph $G'$ induced by $V'$ is connected and $\ttens(G', \latentmat, \conformmat)$, 
computed by the repeated averaging model on $G'$, is minimized.
\end{problem}

\parahead{Algorithms for the {\large {\probSTeam}} problem}
This problem is an NP-hard problem -- as a generalization of the {\sc Set Cover}
problem. However, we can adapt the algorithms we designed for {\probQTeam} to
solve this problem as well. If we denote by $\texttt{Alg}$ an algorithm for solving
{\probQTeam}, then we can use $\texttt{Alg}$ to solve {\probSTeam} by applying
the following two-step procedure:
First, we construct an extended graph $H$ that contains $G=(V,E)$ as well as one
additional node $s$ for every skill $s \in S$. Also, every skill node $s$ is
connected to every node $i$ such that $s \in S_i$. $\texttt{Alg}$ is applied to
solves the {\probQTeam} problem on this extended graph, using the skill nodes
that correspond to skills in $P$ as the seed nodes. 

The result of this first step is a subgraph of $H$ that contains all the skill
nodes required by the project as well as individuals that \emph{cover} all these
skills. One could think that reporting the individuals in this subgraph would
provide an adequate solution. However, this is not the case because removing the
skill nodes from the subgraph might disconnect the subgraph induced by the
individual nodes. For this reason, we need to apply $\texttt{Alg}$ a second time,
now on the original graph $G$ and using as seed nodes the individuals participating
in the solution reported by the first step.

With this strategy, we can directly reuse the algorithms designed for the
{\probQTeam} problem to solve this variant. The running time of such a two-step
procedure will depend on the running time of the algorithm used as a subroutine.




\subsection{Finding low-tension communities with fixed size}
Another variant of the {\probQTeam} problem is one where there are no seed nodes
provided as input but the restriction is put on the size of the output community instead.


Using a construction similar to the one we used for proving Proposition~\ref{prop:hardness},
we can prove that this cardinality variant of our problem is again NP-hard, i.e.\
the problem cannot be solved optimally in polynomial time unless $\text{P}=\text{NP}$.

Although this cardinality version of the problem seems natural, our experiments
demonstrated that it is rather useless in practice. When the value of $k$ is
rather small, the best strategy is to pick one of the many possible subsets of
size $k$ which are rather sparsely connected. Thus, in the absence of guidance
provided in the form of seed members or skills there are a lot of candidate solutions
that are practically equivalent and reporting one of them is not necessarily interesting.

From the algorithmic point of view, the cardinality version of the problem cannot
be solved using some adaptation of the algorithms discussed in Section~\ref{ssec:algorithms}.
For this problem, we obtained the best performance, i.e.\ lowest tension solutions,
using a greedy algorithm that constructs a connected subgraph of cardinality $k$
by repeatedly adding a node that minimally increases the tension of the candidate.


\section{Experiments}
\label{sec:exp}

We now turn to the evaluation of our proposed algorithms for the two problem variants, \probQTeam{} and \probSTeam{}.

\subsection{Experimental setup}
\label{sec:exp_data}
In our setting, each dataset consists of a network together with latent profiles for the nodes. Hence, we first introduce the networks, before explaining our approach for obtaining latent profiles. Then, we present the evaluation measures used throughout our experiments.

\parahead{Networks}
\label{sec:exp_networks} 
Of our collections of networks, two consist of subgraphs extracted from the DBLP co-authorship
database,\!\footnote{\url{http://dblp.uni-trier.de/xml/}}
where vertices represent researchers, and edges represent co-authorship relations.
Specifically, we extracted the ego-nets of radius 2 for some selected high-profile computer
scientists. The resulting ego-nets form the collection denoted as \DBLPed{}.
We also consider the subgraphs induced by researchers who have published in
the ICDM and KDD conferences, respectively, constituting the \DBLPcf{} collection.

Our third collection of networks consists of subgraphs extracted from the Internet Movie Data Base\footnote{\url{http://www.imdb.com}}
where vertices represent actors and edges connect actors who played together in at least one movie.
Specifically, we constructed actor networks from this database by considering some well-known directors and production companies, such as Francis F.\ Coppola or the Warner Bros.\ Entertainment Inc., and extracting the network induced by their movies.  
The resulting networks form the collection denoted as \IMDB{}.

In our problem, we are looking for connected subgraphs. If the seed nodes in \probQTeam{} belong to different components, there is obviously no solution. Hence, in our experiments we consider only the largest component of each of the networks.

The statistics of a sample of the networks from these three collections can be found in Table~\ref{tab:runtimesL}.

\parahead{Profiles}
\label{sec:exp_profiles}
Besides links, the co-authorship and the co-acting networks contain additional information which we exploit to derive structured profiles, as follows.

In the co-authorship networks, we associate nodes, i.e.\ researchers, to the title keywords of papers they authored. We then turn these keywords into profiles by considering the eigenvectors associated to the largest eigenvalues of the incidence matrix of keywords to nodes, scaled to the unit interval.
Vice-versa, we also consider the incidence matrix of researchers to conferences, i.e.\ indications of which author published in which conference and derive another set of profiles by computing the eigenvectors of that matrix.
Neighboring researchers in these networks are collaborators who have published papers together. Hence, they have keywords in common, published in some of the same conferences, and more generally share similar research interest.  Intuitively, they will therefore be assigned similar profile values.

In the co-acting networks, we consider the genres of the movies each actor played in and turn this information into profiles, once again by computing the eigenvectors associated to the largest eigenvalues of the obtained incidence matrices.

\begin{figure}[t]
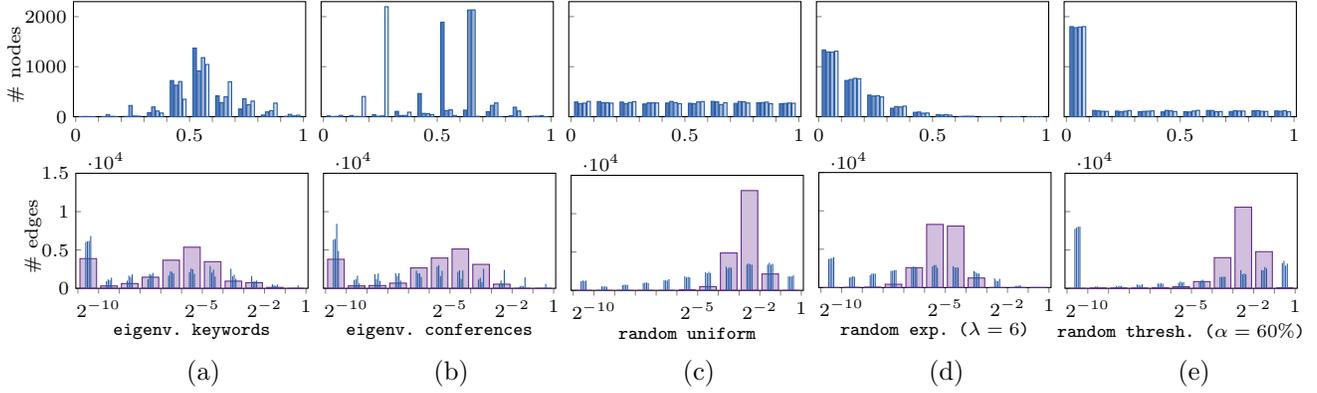

\begin{center}
\hspace{5mm}
\begin{minipage}[t]{0.2\textwidth}
\hplotOpsN{author-author-wpapers.ICDM_all_kw}{xlabel = {}, ylabel = {$\#$ nodes}, ymax=2300}{}
\end{minipage}
\hspace{-4mm}
\begin{minipage}[t]{0.2\textwidth}
\hplotOpsN{author-author-wpapers.ICDM_all_cf}{xlabel = {}, ylabel = {}, yticklabels = {}, ymax=2300}{}
\end{minipage}
\hspace{-4mm}
\begin{minipage}[t]{0.2\textwidth}
\hplotOpsN{author-author-wpapers.ICDM_all_ur}{xlabel = {}, ylabel = {}, yticklabels = {}, ymax=2300}{}
\end{minipage}
\hspace{-4mm}
\begin{minipage}[t]{0.2\textwidth}
\hplotOpsN{author-author-wpapers.ICDM_all_x6r}{xlabel = {}, ylabel = {}, yticklabels = {}, ymax=2300}{}
\end{minipage}
\hspace{-4mm}
\begin{minipage}[t]{0.2\textwidth}
\hplotOpsN{author-author-wpapers.ICDM_all_t6r}{xlabel = {}, ylabel = {}, yticklabels = {}, ymax=2300}{}
\end{minipage}

\hspace{5mm}
\begin{minipage}[t]{0.2\textwidth}
\hplotOpsX{author-author-wpapers.ICDM_all_kw}{xlabel = {\texttt{eigenv.\ keywords}}, ylabel = {$\#$ edges}, xticklabels={, $2^{-10}$, , , , , $2^{-5}$, , , $2^{-2}$, , $1$}, ymax=15000} \\
\centering (a)
\end{minipage}
\hspace{-4mm}
\begin{minipage}[t]{0.2\textwidth}
\hplotOpsX{author-author-wpapers.ICDM_all_cf}{xlabel = {\texttt{eigenv.\ conferences}}, ylabel = {}, xticklabels={, $2^{-10}$, , , , , $2^{-5}$, , , $2^{-2}$, , $1$}, yticklabels = {}, ymax=15000} \\
\centering (b)
\end{minipage}
\hspace{-4mm}
\begin{minipage}[t]{0.2\textwidth}
\hplotOpsX{author-author-wpapers.ICDM_all_ur}{xlabel = {\texttt{random uniform}}, ylabel = {}, xticklabels={, $2^{-10}$, , , , , $2^{-5}$, , , $2^{-2}$, , $1$}, yticklabels = {}, ymax=15000} \\
\centering (c)
\end{minipage}
\hspace{-4mm}
\begin{minipage}[t]{0.2\textwidth}
\hplotOpsX{author-author-wpapers.ICDM_all_x6r}{xlabel = {\texttt{random exp.\ ($\lambda=6$)}}, ylabel = {}, xticklabels={, $2^{-10}$, , , , , $2^{-5}$, , , $2^{-2}$, , $1$}, yticklabels = {}, ymax=15000} \\
\centering (d)
\end{minipage}
\hspace{-4mm}
\begin{minipage}[t]{0.2\textwidth}
\hplotOpsX{author-author-wpapers.ICDM_all_t6r}{xlabel = {\texttt{random thresh.\ ($\alpha=60\%$)}}, ylabel = {}, xticklabels={, $2^{-10}$, , , , , $2^{-5}$, , , $2^{-2}$, , $1$}, yticklabels = {}, ymax=15000} \\
\centering (e)
\end{minipage}
\end{center}
\caption{Distribution of latent profile values, $\innerop_i$, over nodes (top) and of squared weights, $w^2_{ij}$, over edges (bottom) for the \DBLPcf{} \texttt{ICDM} network with various latent profiles.}
\label{fig:hists_opi_dists}
\end{figure}

 The distributions of latent profiles values ($\innerop_i$) over nodes and of squared weights ($w^2_{ij}$) over edges in the \DBLPcf{} \texttt{ICDM} network for latent profiles derived from keywords and from conferences are shown in Figures~\ref{fig:hists_opi_dists}a and~\ref{fig:hists_opi_dists}b respectively.

\renewcommand{\opsdir}{./xps/}

\setlength{\myheight}{0.18\textwidth}

\parahead{Evaluation measures}
\label{sec:exp_eval}

For a solution nodeset $V'$, our main evaluation measure is $\ttens(G(V'), \latentmat, \conformmat)$ --- $\ttens(V')$ for short --- the social tension in the subgraph induced by $V'$ with conformed profiles obtained by applying the repeated averaging process over that subgraph.

Two main properties contribute to a solution subgraph having a low social tension (see Equation~\eqref{eq:overall-tension}). On one hand, finding a subgraph with few edges results in fewer terms in the sum.
On the other hand, finding a subgraph with low tension edges results in small values in the sum. 
Thus, we compute two auxiliary values that provide insight into the nature of the solutions obtained. Namely, for a solution $V'$ we compute the number of edges in the solution, $\abs{E(V')}$, and the average of the squared edge weights in the solution, 
$\avgC{V'} = \frac{1}{\abs{E(V')}} \sum_{(i,j) \in E(V')} w^2_{ij}.$ 

Solutions obtained for different seed sets are hardly comparable. 
Thus, to make the evaluation possible, we standardize the measured values before aggregating them.
Specifically, we use the number of edges in the minimum spanning tree connecting the seed nodes, $\treem$, as a comparison basis (and lower bound) for the number of edges in the solution subgraphs, and divide $\avgC{V'}$ by the corresponding average over the whole graph. 

\newlength{\sjot}
\setlength{\sjot}{\jot}
\setlength{\jot}{0em}

Given a solution $V'$, we take the following evaluation measures, for which lower values are more desirable:
\begin{itemize}
\setlength{\topsep}{0pt}
\setlength{\itemsep}{-1.em}
\setlength{\parsep}{0pt}
 \item [($i$)] the \emph{standardized social tension} (main measure)
\vspace{-.8em} $$ \Mpt = \ttens(V')/ (2 \treem \cdot \avgC{V}),$$
 \item [($ii$)] the \emph{standardized solution size} (auxiliary measure)
\vspace{-.8em} $$ \Mpe = \abs{E(V')}/ \treem, \text{ and}$$
 \item [($iii$)] the \emph{standardized average edge weight} (auxiliary measure)
\vspace{-.8em} $$ \Mpc = \avgC{V'}/\avgC{V}.$$
\end{itemize}

\renewcommand{\minmaxTLHS}{xmin=0.0, xmax=10,}
\renewcommand{\minmaxELHS}{xmin=0.0, xmax=7,}
\renewcommand{\minmaxULHS}{xmin=0.0, xmax=4,}
\renewcommand{\filebasisLHS}{C.Papadimitriou.(1c)}
\renewcommand{\resultdirLHS}{./xps} 

\renewcommand{\minmaxTRHS}{xmin=0.0, xmax=7,}
\renewcommand{\minmaxERHS}{xmin=0.0, xmax=7,}
\renewcommand{\minmaxURHS}{xmin=0.0, xmax=2,}
\renewcommand{\filebasisRHS}{C.Papadimitriou.(xc)}
\renewcommand{\resultdirRHS}{./xps} 

\begin{figure}[t]
\hspace{1.3cm}
\begin{minipage}[t]{\mypwidth}
\begin{tikzpicture}[trim axis left, trim axis right]
\bplotF{\filebasisLHS}{1.0}{3}{\minmaxULHS xlabel={$\Mpt$}, xticklabel pos=right, yticklabels={\algQTree{(e)}, \algQTree{(s)}, \algQPeel{(m)}, \algQPeel{(s)}, \algQPeel{(r)}, \algCock{}}, yticklabel pos=left, xticklabels={}}{\resultdirLHS}
\end{tikzpicture}
\end{minipage}
\begin{minipage}[t]{\mypwidth}
\begin{tikzpicture}[trim axis left, trim axis right]
\bplotF{\filebasisLHS}{1.0}{1}{\minmaxELHS xlabel={$\Mpe$}, xticklabel pos=right, yticklabels={}, xticklabels={}}{\resultdirLHS}
\end{tikzpicture}
\end{minipage}
\begin{minipage}[t]{\mypwidth}
\begin{tikzpicture}[trim axis left, trim axis right]
\bplotF{\filebasisLHS}{1.0}{0}{\minmaxTLHS xlabel={$\Mpc$}, xticklabel pos=right, yticklabel pos=right, yticklabels={}, ylabel={}, xticklabels={}}{\resultdirLHS}
\end{tikzpicture}
\end{minipage}
\hspace{0.55cm}
\begin{minipage}[t]{\mypwidth}
\begin{tikzpicture}[trim axis left, trim axis right]
\bplotF{\filebasisRHS}{1.0}{3}{\minmaxURHS xlabel={$\Mpt$}, xticklabel pos=right, yticklabels={}, yticklabel pos=left, xticklabels={}}{\resultdirRHS}
\end{tikzpicture}
\end{minipage}
\begin{minipage}[t]{\mypwidth}
\begin{tikzpicture}[trim axis left, trim axis right]
\bplotF{\filebasisRHS}{1.0}{1}{\minmaxERHS xlabel={$\Mpe$}, xticklabel pos=right, yticklabels={}, xticklabels={}}{\resultdirRHS}
\end{tikzpicture}
\end{minipage}
\begin{minipage}[t]{\mypwidth}
\begin{tikzpicture}[trim axis left, trim axis right]
\bplotF{\filebasisRHS}{1.0}{0}{\minmaxTRHS xlabel={$\Mpc$}, xticklabel pos=right, yticklabel pos=right, yticklabels={}, ylabel={$D1$}, xticklabels={}}{\resultdirRHS}
\end{tikzpicture}
\end{minipage}

\vspace{0.2em}

\hspace{1.3cm}
\begin{minipage}[t]{\mypwidth}
\begin{tikzpicture}[trim axis left, trim axis right]
\bplotF{\filebasisLHS}{2.0}{3}{\minmaxULHS xticklabel pos=left, yticklabels={\algQTree{(e)}, \algQTree{(s)}, \algQPeel{(m)}, \algQPeel{(s)}, \algQPeel{(r)}, \algCock{}}, yticklabel pos=left, xticklabels={}}{\resultdirLHS}
\end{tikzpicture}
\end{minipage}
\begin{minipage}[t]{\mypwidth}
\begin{tikzpicture}[trim axis left, trim axis right]
\bplotF{\filebasisLHS}{2.0}{1}{\minmaxELHS xticklabel pos=left, yticklabels={}, xticklabels={}}{\resultdirLHS}
\end{tikzpicture}
\end{minipage}
\begin{minipage}[t]{\mypwidth}
\begin{tikzpicture}[trim axis left, trim axis right]
\bplotF{\filebasisLHS}{2.0}{0}{\minmaxTLHS xticklabel pos=left, yticklabel pos=right, yticklabels={}, xticklabels={}, ylabel={}}{\resultdirLHS}
\end{tikzpicture}
\end{minipage}
\hspace{0.55cm}
\begin{minipage}[t]{\mypwidth}
\begin{tikzpicture}[trim axis left, trim axis right]
\bplotF{\filebasisRHS}{2.0}{3}{\minmaxURHS xticklabel pos=left, yticklabels={}, yticklabel pos=left, xticklabels={}}{\resultdirRHS}
\end{tikzpicture}
\end{minipage}
\begin{minipage}[t]{\mypwidth}
\begin{tikzpicture}[trim axis left, trim axis right]
\bplotF{\filebasisRHS}{2.0}{1}{\minmaxERHS xticklabel pos=left, yticklabels={}, xticklabels={}}{\resultdirRHS}
\end{tikzpicture}
\end{minipage}
\begin{minipage}[t]{\mypwidth}
\begin{tikzpicture}[trim axis left, trim axis right]
\bplotF{\filebasisRHS}{2.0}{0}{\minmaxTRHS xticklabel pos=left, yticklabel pos=right, yticklabels={}, xticklabels={}, ylabel={$D2$}}{\resultdirRHS}
\end{tikzpicture}
\end{minipage}

\vspace{-0.4em}

\hspace{1.3cm}
\begin{minipage}[t]{\mypwidth}
\begin{tikzpicture}[trim axis left, trim axis right]
\bplotF{\filebasisLHS}{3.0}{3}{\minmaxULHS xticklabel pos=left, yticklabels={\algQTree{(e)}, \algQTree{(s)}, \algQPeel{(m)}, \algQPeel{(s)}, \algQPeel{(r)}, \algCock{}}, yticklabel pos=left}{\resultdirLHS}
\end{tikzpicture}
\end{minipage}
\begin{minipage}[t]{\mypwidth}
\begin{tikzpicture}[trim axis left, trim axis right]
\bplotF{\filebasisLHS}{3.0}{1}{\minmaxELHS xticklabel pos=left, yticklabels={}}{\resultdirLHS}
\end{tikzpicture}
\end{minipage}
\begin{minipage}[t]{\mypwidth}
\begin{tikzpicture}[trim axis left, trim axis right]
\bplotF{\filebasisLHS}{3.0}{0}{\minmaxTLHS xticklabel pos=left, yticklabel pos=right, yticklabels={}, ylabel={}}{\resultdirLHS}
\end{tikzpicture}
\end{minipage}
\hspace{0.55cm}
\begin{minipage}[t]{\mypwidth}
\begin{tikzpicture}[trim axis left, trim axis right]
\bplotF{\filebasisRHS}{3.0}{3}{\minmaxURHS xticklabel pos=left, yticklabels={}, yticklabel pos=left}{\resultdirRHS}
\end{tikzpicture}
\end{minipage}
\begin{minipage}[t]{\mypwidth}
\begin{tikzpicture}[trim axis left, trim axis right]
\bplotF{\filebasisRHS}{3.0}{1}{\minmaxERHS xticklabel pos=left, yticklabels={}}{\resultdirRHS}
\end{tikzpicture}
\end{minipage}
\begin{minipage}[t]{\mypwidth}
\begin{tikzpicture}[trim axis left, trim axis right]
\bplotF{\filebasisRHS}{3.0}{0}{\minmaxTRHS xticklabel pos=left, yticklabel pos=right, yticklabels={}, ylabel={$D3$}}{\resultdirRHS}
\end{tikzpicture}
\end{minipage}

\caption{Results for the \probQTeam{} problem on the \texttt{C.Papadimitriou} with single-attribute (left) and ten-attribute (right) latent profiles derived from conferences.}
\label{fig:hists_opiQ_papadimitriou}
\end{figure}
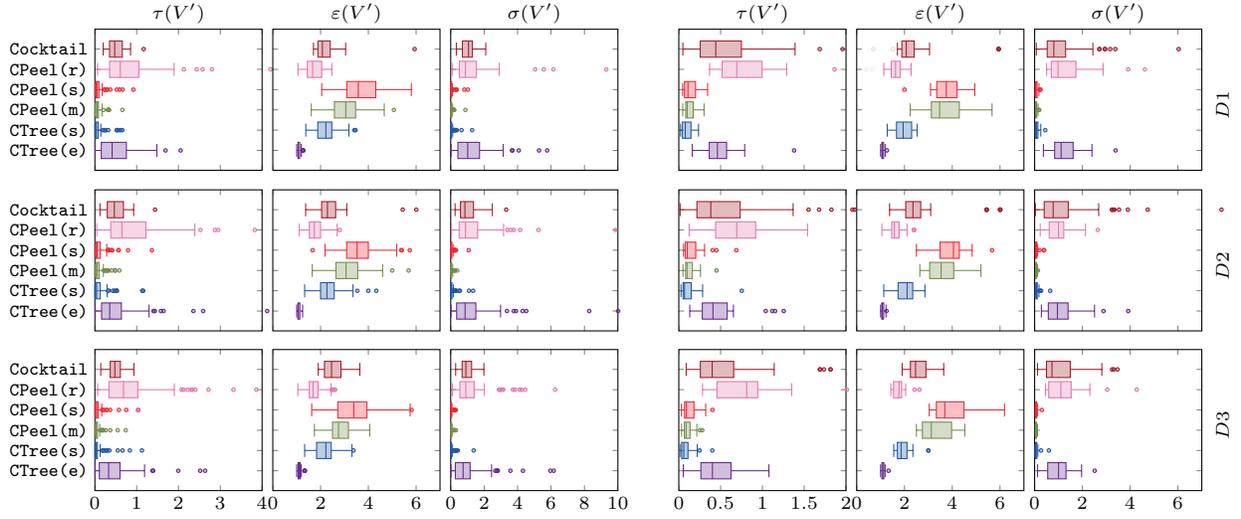

\subsection{Community search with seed nodes}
We now compare the different variants of our proposed algorithms, \algQTree{} and
\algQPeel{}, for solving the \probQTeam{} problem (see
Section~\ref{sec:qproblem}).

\parahead{Generating sets of seed nodes}
\label{sec:exp_seeds}
For each dataset, i.e.\ each pair of network and latent profiles, we run
each algorithm with a number of different sets of seed nodes $Q$. Here
we restrict ourselves to sets of seven and four seed nodes, i.e.\ $\abs{Q}=7$ and $\abs{Q}=4$, for the co-authorship networks and the co-acting networks respectively,
as representative scenarios for the community-search problem.

As we expect the distance between the nodes in the seed set to
have an impact on the behavior of the algorithms, we want to sample
seed sets across the range of possible distances and to group them
based on this criterion. Thus, we generate a thousand seed sets
and look at the maximum pairwise distance within each set. We then
select at most $30$ seed sets from the 
10-33\%,
33-66\%, and
66-90\%
percentiles of the distance
distribution. The resulting three groups of seed sets are denoted as
$D1$, $D2$ and $D3$, from tight seed sets to more
dispersed ones. Results in 
Figure~\ref{fig:hists_opiQ_papadimitriou} 
are presented aggregated according to these distance groups.

\parahead{Single-attribute and multi-attribute profiles}
For each latent profile construction scheme, i.e.\ whether derived from keywords, conferences or movie genres (see Section~\ref{sec:exp_profiles}), we can either consider the column vectors separately, thereby obtaining several single-attribute profiles, where each node is associated to a single profile value, or consider the entire matrix at once thereby obtaining one multi-attribute profile, where each node is associated to a multi-attribute profile vector.

In our experiments, we take the first four columns of the matrix as four separate single-attribute profiles and the entire matrix as one multi-attribute profile (6 and 21 attributes for the \DBLP{} and \IMDB{} datasets, respectively). 
Results for the \DBLPed{} \texttt{C.Papadimitriou} for single-attribute (left) and multi-attribute (right) latent profiles derived from conferences are shown in
Figure~\ref{fig:hists_opiQ_papadimitriou}. 
Similarly, results for the \IMDB{} \texttt{F.F.Coppola} network for single-attribute (left) and multi-attribute (right) latent profiles derived from movie genres are shown in Figure~\ref{fig:hists_opiQ_coppola}.

\renewcommand{\minmaxTLHS}{xmin=0.0, xmax=4,}
\renewcommand{\minmaxELHS}{xmin=0.0, xmax=8,}
\renewcommand{\minmaxULHS}{xmin=0.0, xmax=1.75,}
\renewcommand{\filebasisLHS}{Coppola.(1l)}
\renewcommand{\resultdirLHS}{./xps} 

\renewcommand{\minmaxTRHS}{xmin=0.0, xmax=4,}
\renewcommand{\minmaxERHS}{xmin=0.0, xmax=8,}
\renewcommand{\minmaxURHS}{xmin=0.0, xmax=1.75,}
\renewcommand{\filebasisRHS}{Coppola.(xl)}
\renewcommand{\resultdirRHS}{./xps} 

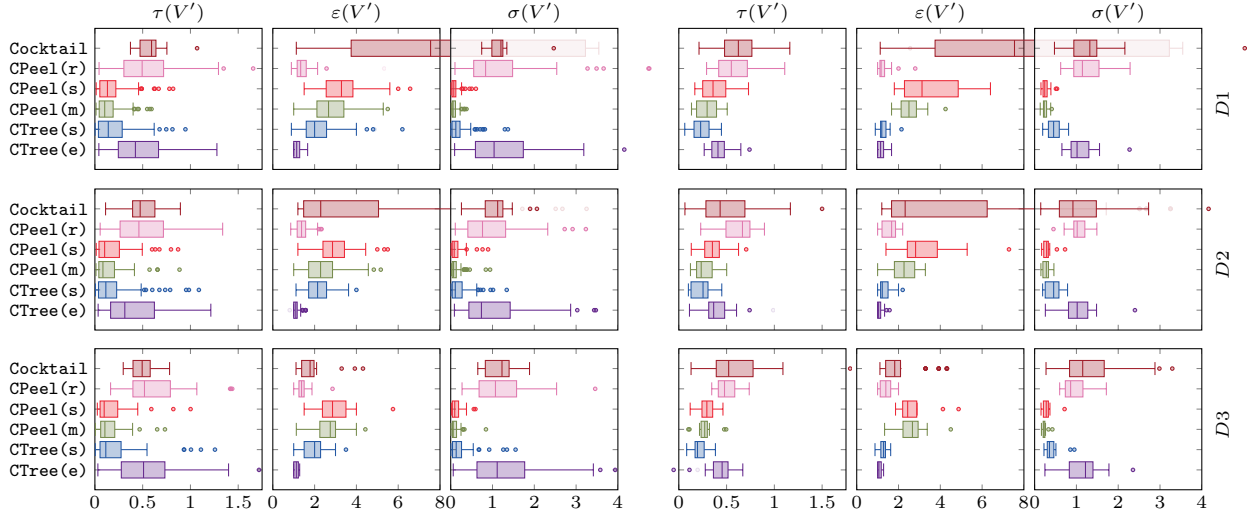
\begin{figure}[t]
\hspace{1.3cm}
\begin{minipage}[t]{\mypwidth}
\begin{tikzpicture}[trim axis left, trim axis right]
\bplotF{\filebasisLHS}{1.0}{3}{\minmaxULHS xlabel={$\Mpt$}, xticklabel pos=right, yticklabels={\algQTree{(e)}, \algQTree{(s)}, \algQPeel{(m)}, \algQPeel{(s)}, \algQPeel{(r)}, \algCock{}}, yticklabel pos=left, xticklabels={}}{\resultdirLHS}
\end{tikzpicture}
\end{minipage}
\begin{minipage}[t]{\mypwidth}
\begin{tikzpicture}[trim axis left, trim axis right]
\bplotF{\filebasisLHS}{1.0}{1}{\minmaxELHS xlabel={$\Mpe$}, xticklabel pos=right, yticklabels={}, xticklabels={}}{\resultdirLHS}
\end{tikzpicture}
\end{minipage}
\begin{minipage}[t]{\mypwidth}
\begin{tikzpicture}[trim axis left, trim axis right]
\bplotF{\filebasisLHS}{1.0}{0}{\minmaxTLHS xlabel={$\Mpc$}, xticklabel pos=right, yticklabel pos=right, yticklabels={}, ylabel={}, xticklabels={}}{\resultdirLHS}
\end{tikzpicture}
\end{minipage}
\hspace{0.55cm}
\begin{minipage}[t]{\mypwidth}
\begin{tikzpicture}[trim axis left, trim axis right]
\bplotF{\filebasisRHS}{1.0}{3}{\minmaxURHS xlabel={$\Mpt$}, xticklabel pos=right, yticklabels={}, yticklabel pos=left, xticklabels={}}{\resultdirRHS}
\end{tikzpicture}
\end{minipage}
\begin{minipage}[t]{\mypwidth}
\begin{tikzpicture}[trim axis left, trim axis right]
\bplotF{\filebasisRHS}{1.0}{1}{\minmaxERHS xlabel={$\Mpe$}, xticklabel pos=right, yticklabels={}, xticklabels={}}{\resultdirRHS}
\end{tikzpicture}
\end{minipage}
\begin{minipage}[t]{\mypwidth}
\begin{tikzpicture}[trim axis left, trim axis right]
\bplotF{\filebasisRHS}{1.0}{0}{\minmaxTRHS xlabel={$\Mpc$}, xticklabel pos=right, yticklabel pos=right, yticklabels={}, ylabel={$D1$}, xticklabels={}}{\resultdirRHS}
\end{tikzpicture}
\end{minipage}

\vspace{0.2em}

\hspace{1.3cm}
\begin{minipage}[t]{\mypwidth}
\begin{tikzpicture}[trim axis left, trim axis right]
\bplotF{\filebasisLHS}{2.0}{3}{\minmaxULHS xticklabel pos=left, yticklabels={\algQTree{(e)}, \algQTree{(s)}, \algQPeel{(m)}, \algQPeel{(s)}, \algQPeel{(r)}, \algCock{}}, yticklabel pos=left, xticklabels={}}{\resultdirLHS}
\end{tikzpicture}
\end{minipage}
\begin{minipage}[t]{\mypwidth}
\begin{tikzpicture}[trim axis left, trim axis right]
\bplotF{\filebasisLHS}{2.0}{1}{\minmaxELHS xticklabel pos=left, yticklabels={}, xticklabels={}}{\resultdirLHS}
\end{tikzpicture}
\end{minipage}
\begin{minipage}[t]{\mypwidth}
\begin{tikzpicture}[trim axis left, trim axis right]
\bplotF{\filebasisLHS}{2.0}{0}{\minmaxTLHS xticklabel pos=left, yticklabel pos=right, yticklabels={}, xticklabels={}, ylabel={}}{\resultdirLHS}
\end{tikzpicture}
\end{minipage}
\hspace{0.55cm}
\begin{minipage}[t]{\mypwidth}
\begin{tikzpicture}[trim axis left, trim axis right]
\bplotF{\filebasisRHS}{2.0}{3}{\minmaxURHS xticklabel pos=left, yticklabels={}, yticklabel pos=left, xticklabels={}}{\resultdirRHS}
\end{tikzpicture}
\end{minipage}
\begin{minipage}[t]{\mypwidth}
\begin{tikzpicture}[trim axis left, trim axis right]
\bplotF{\filebasisRHS}{2.0}{1}{\minmaxERHS xticklabel pos=left, yticklabels={}, xticklabels={}}{\resultdirRHS}
\end{tikzpicture}
\end{minipage}
\begin{minipage}[t]{\mypwidth}
\begin{tikzpicture}[trim axis left, trim axis right]
\bplotF{\filebasisRHS}{2.0}{0}{\minmaxTRHS xticklabel pos=left, yticklabel pos=right, yticklabels={}, xticklabels={}, ylabel={$D2$}}{\resultdirRHS}
\end{tikzpicture}
\end{minipage}

\vspace{-0.4em}

\hspace{1.3cm}
\begin{minipage}[t]{\mypwidth}
\begin{tikzpicture}[trim axis left, trim axis right]
\bplotF{\filebasisLHS}{3.0}{3}{\minmaxULHS xticklabel pos=left, yticklabels={\algQTree{(e)}, \algQTree{(s)}, \algQPeel{(m)}, \algQPeel{(s)}, \algQPeel{(r)}, \algCock{}}, yticklabel pos=left}{\resultdirLHS}
\end{tikzpicture}
\end{minipage}
\begin{minipage}[t]{\mypwidth}
\begin{tikzpicture}[trim axis left, trim axis right]
\bplotF{\filebasisLHS}{3.0}{1}{\minmaxELHS xticklabel pos=left, yticklabels={}}{\resultdirLHS}
\end{tikzpicture}
\end{minipage}
\begin{minipage}[t]{\mypwidth}
\begin{tikzpicture}[trim axis left, trim axis right]
\bplotF{\filebasisLHS}{3.0}{0}{\minmaxTLHS xticklabel pos=left, yticklabel pos=right, yticklabels={}, ylabel={}}{\resultdirLHS}
\end{tikzpicture}
\end{minipage}
\hspace{0.55cm}
\begin{minipage}[t]{\mypwidth}
\begin{tikzpicture}[trim axis left, trim axis right]
\bplotF{\filebasisRHS}{3.0}{3}{\minmaxURHS xticklabel pos=left, yticklabels={}, yticklabel pos=left}{\resultdirRHS}
\end{tikzpicture}
\end{minipage}
\begin{minipage}[t]{\mypwidth}
\begin{tikzpicture}[trim axis left, trim axis right]
\bplotF{\filebasisRHS}{3.0}{1}{\minmaxERHS xticklabel pos=left, yticklabels={}}{\resultdirRHS}
\end{tikzpicture}
\end{minipage}
\begin{minipage}[t]{\mypwidth}
\begin{tikzpicture}[trim axis left, trim axis right]
\bplotF{\filebasisRHS}{3.0}{0}{\minmaxTRHS xticklabel pos=left, yticklabel pos=right, yticklabels={}, ylabel={$D3$}}{\resultdirRHS}
\end{tikzpicture}
\end{minipage}

\caption{Results for the \probQTeam{} problem on the \IMDB{} \texttt{F.F.Coppola} with single-attribute (left) and ten-attribute (right) latent profiles derived from movie genres.}
\label{fig:hists_opiQ_coppola}
\end{figure}

We observe that the algorithms that exploit the profiles of individuals
outperform the other two variants in (almost) all cases. 
Recall that neither \algQTree{(e)} nor \algQPeel{(r)} consider the
profiles of individuals. All they can do is minimize the social
tension by finding a small subgraph to connect the seed nodes. Our
results show that they are indeed quite effective at minimizing the
number of edges in the reported solutions, typically achieving the
lowest values of $\Mpe$ (middle column in each block of
Figure~\ref{fig:hists_opiQ_papadimitriou} and~\ref{fig:hists_opiQ_coppola}). However,
\algQTree{(s)}, \algQPeel{(m)}, and \algQPeel{(s)}, the profile-aware
variants, find solutions with lower edge weights, i.e.\ achieving
lower values for $\Mpc$ (right hand side column), at the cost of
including extra edges. This gives them an advantage for
minimizing social tension, obtaining lower values for $\Mpt$
(left hand side column).

This pattern clearly holds whether we consider single-attribute or multi-attribute profiles.

\parahead{Comparison to finding dense communities}
In addition to our proposed algorithms, we also obtain communities by applying the \texttt{GreedyFast} algorithm of Sozio and Gionis~\ccite{DBLP:conf/kdd/SozioG10}, denoted here as \algCock{}, on the networks with the same sets of seed individuals. The aim of this algorithm is to find a subgraph that connects the seeds while maximizing the minimum degree among selected nodes. This algorithm considers neither profiles nor social tension; Nevertheless, we can compute the tension of the community that consists of the nodes returned as  a solution and compare it to the communities obtained with our algorithms.
Furthermore, \algCock{} requires the user to set a value for the upper bound on the size of the solution. We set this value to $k=200$, as it seems to result in reasonable runtimes while allowing the algorithm to construct a solution in most cases. The cases where the algorithm fails to return a solution are left out from our statistics.

On the \DBLPed{} \texttt{C.Papadimitriou} network the solutions returned by \algCock{} are comparable in size to those of our \algQTree{(s)} algorithm (middle column in each block of
Figure~\ref{fig:hists_opiQ_papadimitriou}). The quality of edges selected in its solutions is on par with \algQTree{(e)} and \algQPeel{(r)} (right hand side column), which are also oblivious to the profiles of individuals.
On the \IMDB{} \texttt{F.F.Coppola} network, the solutions returned by \algCock{} tend to be much larger than any of our algorithms (middle column in each block of Figure~\ref{fig:hists_opiQ_coppola}).
Expectedly, \algCock{} appears poorly suited for the task of finding low-tension communities.

\parahead{Impact of the profiles distribution}
We observed that the \algQTree{(s)} variant appears to typically achieve the best
performance, followed closely by \algQPeel{(m)}. Yet, variations can be
observed depending on the network structure and on the properties of the
opinions distribution. 

Thus, we next investigate the impact of the distribution of profile values on the behavior of the algorithms.
In order to do so, we construct random latent profiles under different sampling distributions.
More specifically, beside the \emph{uniform distribution}, we either sample from the \emph{exponential distribution} with parameter $\lambda$, or we set the profile of a fraction $\alpha$ of the nodes to zero, while the rest is sampled uniformly at random, obtaining a \emph{thresholded distribution}. 

For this experiment, we focus on the \DBLPcf{} \texttt{ICDM} network and single-attribute latent profiles. 

The distributions of latent profiles values ($\innerop_i$) over nodes and of squared weights ($w^2_{ij}$) over edges in this network for various latent profiles are shown in Figure~\ref{fig:hists_opi_dists}.
Latent profiles were sampled following each of the three random distributions, in addition to considering latent profiles derived from the keywords and from the conferences respectively. The narrow blue bars show the distributions for the first four single-attribute profiles considered separately, while the broader purple bars show the distributions for the ten-attribute profiles resulting from the combination of the respective single-attribute profiles.

The results obtained for these different profile construction schemes, each time applying the algorithms to the first four single-attribute profiles considered separately are shown in Figure~\ref{fig:hists_opiQ_six}.

\renewcommand{\minmaxTLHS}{xmin=0.0, xmax=5,}
\renewcommand{\minmaxELHS}{xmin=0.0, xmax=6.25,}
\renewcommand{\minmaxULHS}{xmin=0.0, xmax=1.75,}
\renewcommand{\filebasisLHS}{ICDM(cf)}
\renewcommand{\resultdirLHS}{./xps} 

\renewcommand{\minmaxTRHS}{xmin=0.0, xmax=5,}
\renewcommand{\minmaxERHS}{xmin=0.0, xmax=6.25,}
\renewcommand{\minmaxURHS}{xmin=0.0, xmax=1.75,}
\renewcommand{\filebasisRHS}{ICDM(kw)}
\renewcommand{\resultdirRHS}{./xps} 

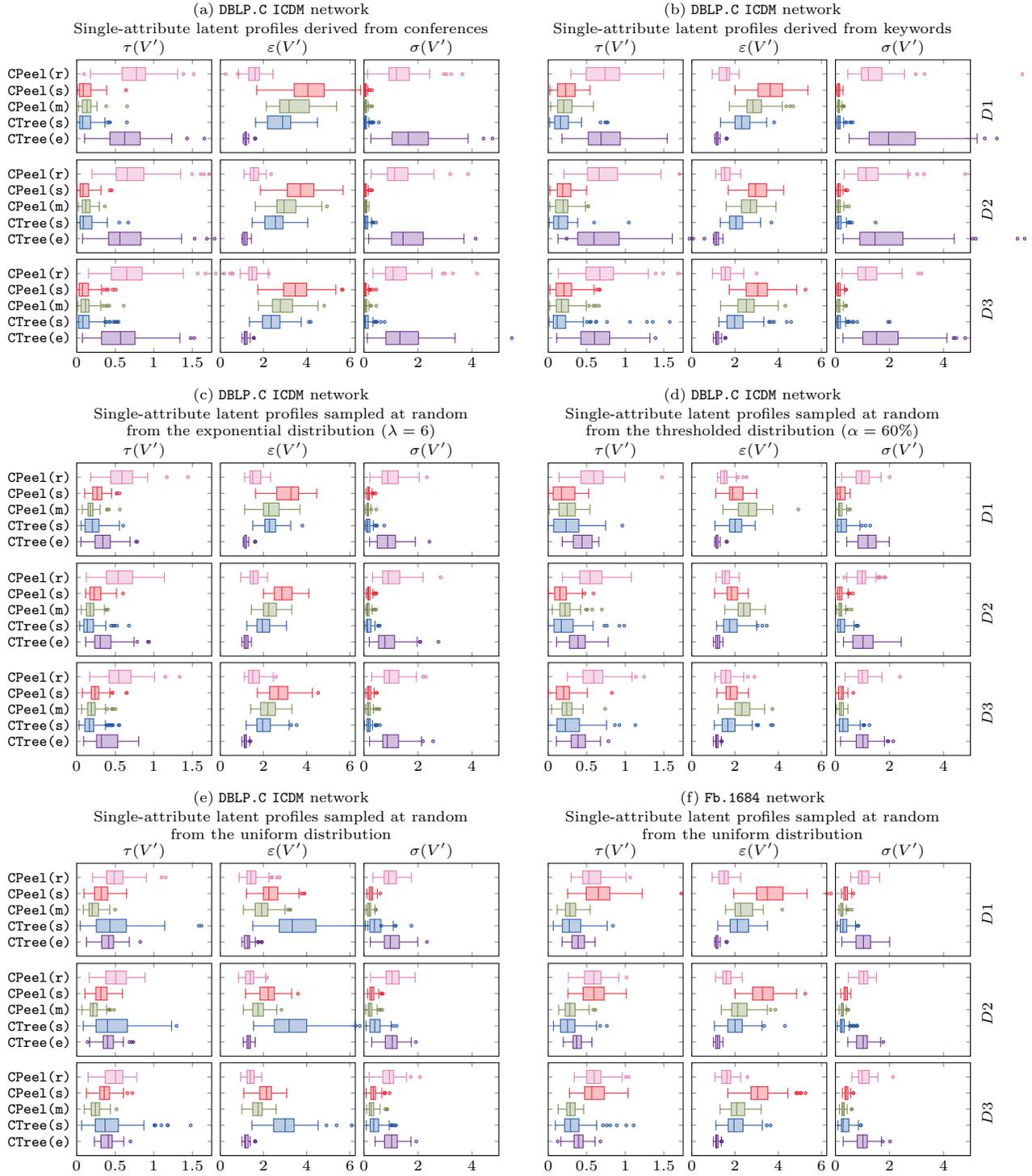
\begin{figure}[p]
\hspace{1.3cm}
\begin{minipage}[t]{\mypwidth}
\begin{tikzpicture}[trim axis left, trim axis right]
\bplotR{\filebasisLHS}{1.0}{3}{\minmaxULHS xlabel={$\Mpt$}, xticklabel pos=right, yticklabels={\algQTree{(e)}, \algQTree{(s)}, \algQPeel{(m)}, \algQPeel{(s)}, \algQPeel{(r)}}, yticklabel pos=left, xticklabels={}}{\resultdirLHS}
\end{tikzpicture}
\end{minipage}
\begin{minipage}[t]{\mypwidth}
\begin{tikzpicture}[trim axis left, trim axis right]
\draw (1,2.1) node[above, font=\scriptsize]{(a)\;\DBLPcf{} \texttt{ICDM} network};
\draw (1,1.8) node[above, font=\scriptsize]{Single-attribute latent profiles derived from conferences};
\bplotR{\filebasisLHS}{1.0}{1}{\minmaxELHS xlabel={$\Mpe$}, xticklabel pos=right, yticklabels={}, xticklabels={}}{\resultdirLHS}
\end{tikzpicture}
\end{minipage}
\begin{minipage}[t]{\mypwidth}
\begin{tikzpicture}[trim axis left, trim axis right]
\bplotR{\filebasisLHS}{1.0}{0}{\minmaxTLHS xlabel={$\Mpc$}, xticklabel pos=right, yticklabel pos=right, yticklabels={}, ylabel={}, xticklabels={}}{\resultdirLHS}
\end{tikzpicture}
\end{minipage}
\hspace{0.55cm}
\begin{minipage}[t]{\mypwidth}
\begin{tikzpicture}[trim axis left, trim axis right]
\bplotR{\filebasisRHS}{1.0}{3}{\minmaxURHS xlabel={$\Mpt$}, xticklabel pos=right, yticklabels={}, yticklabel pos=left, xticklabels={}}{\resultdirRHS}
\end{tikzpicture}
\end{minipage}
\begin{minipage}[t]{\mypwidth}
\begin{tikzpicture}[trim axis left, trim axis right]
\draw (1,2.1) node[above, font=\scriptsize]{(b)\;\DBLPcf{} \texttt{ICDM} network};
\draw (1,1.8) node[above, font=\scriptsize]{Single-attribute latent profiles derived from keywords};
\bplotR{\filebasisRHS}{1.0}{1}{\minmaxERHS xlabel={$\Mpe$}, xticklabel pos=right, yticklabels={}, xticklabels={}}{\resultdirRHS}
\end{tikzpicture}
\end{minipage}
\begin{minipage}[t]{\mypwidth}
\begin{tikzpicture}[trim axis left, trim axis right]
\bplotR{\filebasisRHS}{1.0}{0}{\minmaxTRHS xlabel={$\Mpc$}, xticklabel pos=right, yticklabel pos=right, yticklabels={}, ylabel={$D1$}, xticklabels={}}{\resultdirRHS}
\end{tikzpicture}
\end{minipage}

\vspace{0.2em}

\hspace{1.3cm}
\begin{minipage}[t]{\mypwidth}
\begin{tikzpicture}[trim axis left, trim axis right]
\bplotR{\filebasisLHS}{2.0}{3}{\minmaxULHS xticklabel pos=left, yticklabels={\algQTree{(e)}, \algQTree{(s)}, \algQPeel{(m)}, \algQPeel{(s)}, \algQPeel{(r)}}, yticklabel pos=left, xticklabels={}}{\resultdirLHS}
\end{tikzpicture}
\end{minipage}
\begin{minipage}[t]{\mypwidth}
\begin{tikzpicture}[trim axis left, trim axis right]
\bplotR{\filebasisLHS}{2.0}{1}{\minmaxELHS xticklabel pos=left, yticklabels={}, xticklabels={}}{\resultdirLHS}
\end{tikzpicture}
\end{minipage}
\begin{minipage}[t]{\mypwidth}
\begin{tikzpicture}[trim axis left, trim axis right]
\bplotR{\filebasisLHS}{2.0}{0}{\minmaxTLHS xticklabel pos=left, yticklabel pos=right, yticklabels={}, xticklabels={}, ylabel={}}{\resultdirLHS}
\end{tikzpicture}
\end{minipage}
\hspace{0.55cm}
\begin{minipage}[t]{\mypwidth}
\begin{tikzpicture}[trim axis left, trim axis right]
\bplotR{\filebasisRHS}{2.0}{3}{\minmaxURHS xticklabel pos=left, yticklabels={}, yticklabel pos=left, xticklabels={}}{\resultdirRHS}
\end{tikzpicture}
\end{minipage}
\begin{minipage}[t]{\mypwidth}
\begin{tikzpicture}[trim axis left, trim axis right]
\bplotR{\filebasisRHS}{2.0}{1}{\minmaxERHS xticklabel pos=left, yticklabels={}, xticklabels={}}{\resultdirRHS}
\end{tikzpicture}
\end{minipage}
\begin{minipage}[t]{\mypwidth}
\begin{tikzpicture}[trim axis left, trim axis right]
\bplotR{\filebasisRHS}{2.0}{0}{\minmaxTRHS xticklabel pos=left, yticklabel pos=right, yticklabels={}, xticklabels={}, ylabel={$D2$}}{\resultdirRHS}
\end{tikzpicture}
\end{minipage}

\vspace{-0.4em}

\hspace{1.3cm}
\begin{minipage}[t]{\mypwidth}
\begin{tikzpicture}[trim axis left, trim axis right]
\bplotR{\filebasisLHS}{3.0}{3}{\minmaxULHS xticklabel pos=left, yticklabels={\algQTree{(e)}, \algQTree{(s)}, \algQPeel{(m)}, \algQPeel{(s)}, \algQPeel{(r)}}, yticklabel pos=left}{\resultdirLHS}
\end{tikzpicture}
\end{minipage}
\begin{minipage}[t]{\mypwidth}
\begin{tikzpicture}[trim axis left, trim axis right]
\bplotR{\filebasisLHS}{3.0}{1}{\minmaxELHS xticklabel pos=left, yticklabels={}}{\resultdirLHS}
\end{tikzpicture}
\end{minipage}
\begin{minipage}[t]{\mypwidth}
\begin{tikzpicture}[trim axis left, trim axis right]
\bplotR{\filebasisLHS}{3.0}{0}{\minmaxTLHS xticklabel pos=left, yticklabel pos=right, yticklabels={}, ylabel={}}{\resultdirLHS}
\end{tikzpicture}
\end{minipage}
\hspace{0.55cm}
\begin{minipage}[t]{\mypwidth}
\begin{tikzpicture}[trim axis left, trim axis right]
\bplotR{\filebasisRHS}{3.0}{3}{\minmaxURHS xticklabel pos=left, yticklabels={}, yticklabel pos=left}{\resultdirRHS}
\end{tikzpicture}
\end{minipage}
\begin{minipage}[t]{\mypwidth}
\begin{tikzpicture}[trim axis left, trim axis right]
\bplotR{\filebasisRHS}{3.0}{1}{\minmaxERHS xticklabel pos=left, yticklabels={}}{\resultdirRHS}
\end{tikzpicture}
\end{minipage}
\begin{minipage}[t]{\mypwidth}
\begin{tikzpicture}[trim axis left, trim axis right]
\bplotR{\filebasisRHS}{3.0}{0}{\minmaxTRHS xticklabel pos=left, yticklabel pos=right, yticklabels={}, ylabel={$D3$}}{\resultdirRHS}
\end{tikzpicture}
\end{minipage}

\vspace{0.2em}

\renewcommand{\filebasisLHS}{ICDM(x6r)}
\renewcommand{\filebasisRHS}{ICDM(t6r)}

\hspace{1.3cm}
\begin{minipage}[t]{\mypwidth}
\begin{tikzpicture}[trim axis left, trim axis right]
\bplotR{\filebasisLHS}{1.0}{3}{\minmaxULHS xlabel={$\Mpt$}, xticklabel pos=right, yticklabels={\algQTree{(e)}, \algQTree{(s)}, \algQPeel{(m)}, \algQPeel{(s)}, \algQPeel{(r)}}, yticklabel pos=left, xticklabels={}}{\resultdirLHS}
\end{tikzpicture}
\end{minipage}
\begin{minipage}[t]{\mypwidth}
\begin{tikzpicture}[trim axis left, trim axis right]
\draw (1,2.4) node[above, font=\scriptsize]{(c)\;\DBLPcf{} \texttt{ICDM} network};
\draw (1,2.1) node[above, font=\scriptsize]{Single-attribute latent profiles sampled at random};
\draw (1,1.8) node[above, font=\scriptsize]{from the exponential distribution ($\lambda=6$)};
\bplotR{\filebasisLHS}{1.0}{1}{\minmaxELHS xlabel={$\Mpe$}, xticklabel pos=right, yticklabels={}, xticklabels={}}{\resultdirLHS}
\end{tikzpicture}
\end{minipage}
\begin{minipage}[t]{\mypwidth}
\begin{tikzpicture}[trim axis left, trim axis right]
\bplotR{\filebasisLHS}{1.0}{0}{\minmaxTLHS xlabel={$\Mpc$}, xticklabel pos=right, yticklabel pos=right, yticklabels={}, ylabel={}, xticklabels={}}{\resultdirLHS}
\end{tikzpicture}
\end{minipage}
\hspace{0.55cm}
\begin{minipage}[t]{\mypwidth}
\begin{tikzpicture}[trim axis left, trim axis right]
\bplotR{\filebasisRHS}{1.0}{3}{\minmaxURHS xlabel={$\Mpt$}, xticklabel pos=right, yticklabels={}, yticklabel pos=left, xticklabels={}}{\resultdirRHS}
\end{tikzpicture}
\end{minipage}
\begin{minipage}[t]{\mypwidth}
\begin{tikzpicture}[trim axis left, trim axis right]
\draw (1,2.4) node[above, font=\scriptsize]{(d)\;\DBLPcf{} \texttt{ICDM} network};
\draw (1,2.1) node[above, font=\scriptsize]{Single-attribute latent profiles sampled at random};
\draw (1,1.8) node[above, font=\scriptsize]{from the thresholded distribution ($\alpha=60\%$)};
\bplotR{\filebasisRHS}{1.0}{1}{\minmaxERHS xlabel={$\Mpe$}, xticklabel pos=right, yticklabels={}, xticklabels={}}{\resultdirRHS}
\end{tikzpicture}
\end{minipage}
\begin{minipage}[t]{\mypwidth}
\begin{tikzpicture}[trim axis left, trim axis right]
\bplotR{\filebasisRHS}{1.0}{0}{\minmaxTRHS xlabel={$\Mpc$}, xticklabel pos=right, yticklabel pos=right, yticklabels={}, ylabel={$D1$}, xticklabels={}}{\resultdirRHS}
\end{tikzpicture}
\end{minipage}

\vspace{.2em}

\hspace{1.3cm}
\begin{minipage}[t]{\mypwidth}
\begin{tikzpicture}[trim axis left, trim axis right]
\bplotR{\filebasisLHS}{2.0}{3}{\minmaxULHS xticklabel pos=left, yticklabels={\algQTree{(e)}, \algQTree{(s)}, \algQPeel{(m)}, \algQPeel{(s)}, \algQPeel{(r)}}, yticklabel pos=left, xticklabels={}}{\resultdirLHS}
\end{tikzpicture}
\end{minipage}
\begin{minipage}[t]{\mypwidth}
\begin{tikzpicture}[trim axis left, trim axis right]
\bplotR{\filebasisLHS}{2.0}{1}{\minmaxELHS xticklabel pos=left, yticklabels={}, xticklabels={}}{\resultdirLHS}
\end{tikzpicture}
\end{minipage}
\begin{minipage}[t]{\mypwidth}
\begin{tikzpicture}[trim axis left, trim axis right]
\bplotR{\filebasisLHS}{2.0}{0}{\minmaxTLHS xticklabel pos=left, yticklabel pos=right, yticklabels={}, xticklabels={}, ylabel={}}{\resultdirLHS}
\end{tikzpicture}
\end{minipage}
\hspace{0.55cm}
\begin{minipage}[t]{\mypwidth}
\begin{tikzpicture}[trim axis left, trim axis right]
\bplotR{\filebasisRHS}{2.0}{3}{\minmaxURHS xticklabel pos=left, yticklabels={}, yticklabel pos=left, xticklabels={}}{\resultdirRHS}
\end{tikzpicture}
\end{minipage}
\begin{minipage}[t]{\mypwidth}
\begin{tikzpicture}[trim axis left, trim axis right]
\bplotR{\filebasisRHS}{2.0}{1}{\minmaxERHS xticklabel pos=left, yticklabels={}, xticklabels={}}{\resultdirRHS}
\end{tikzpicture}
\end{minipage}
\begin{minipage}[t]{\mypwidth}
\begin{tikzpicture}[trim axis left, trim axis right]
\bplotR{\filebasisRHS}{2.0}{0}{\minmaxTRHS xticklabel pos=left, yticklabel pos=right, yticklabels={}, xticklabels={}, ylabel={$D2$}}{\resultdirRHS}
\end{tikzpicture}
\end{minipage}

\vspace{-0.4em}

\hspace{1.3cm}
\begin{minipage}[t]{\mypwidth}
\begin{tikzpicture}[trim axis left, trim axis right]
\bplotR{\filebasisLHS}{3.0}{3}{\minmaxULHS xticklabel pos=left, yticklabels={\algQTree{(e)}, \algQTree{(s)}, \algQPeel{(m)}, \algQPeel{(s)}, \algQPeel{(r)}}, yticklabel pos=left}{\resultdirLHS}
\end{tikzpicture}
\end{minipage}
\begin{minipage}[t]{\mypwidth}
\begin{tikzpicture}[trim axis left, trim axis right]
\bplotR{\filebasisLHS}{3.0}{1}{\minmaxELHS xticklabel pos=left, yticklabels={}}{\resultdirLHS}
\end{tikzpicture}
\end{minipage}
\begin{minipage}[t]{\mypwidth}
\begin{tikzpicture}[trim axis left, trim axis right]
\bplotR{\filebasisLHS}{3.0}{0}{\minmaxTLHS xticklabel pos=left, yticklabel pos=right, yticklabels={}, ylabel={}}{\resultdirLHS}
\end{tikzpicture}
\end{minipage}
\hspace{0.55cm}
\begin{minipage}[t]{\mypwidth}
\begin{tikzpicture}[trim axis left, trim axis right]
\bplotR{\filebasisRHS}{3.0}{3}{\minmaxURHS xticklabel pos=left, yticklabels={}, yticklabel pos=left}{\resultdirRHS}
\end{tikzpicture}
\end{minipage}
\begin{minipage}[t]{\mypwidth}
\begin{tikzpicture}[trim axis left, trim axis right]
\bplotR{\filebasisRHS}{3.0}{1}{\minmaxERHS xticklabel pos=left, yticklabels={}}{\resultdirRHS}
\end{tikzpicture}
\end{minipage}
\begin{minipage}[t]{\mypwidth}
\begin{tikzpicture}[trim axis left, trim axis right]
\bplotR{\filebasisRHS}{3.0}{0}{\minmaxTRHS xticklabel pos=left, yticklabel pos=right, yticklabels={}, ylabel={$D3$}}{\resultdirRHS}
\end{tikzpicture}
\end{minipage}

\vspace{0.2em}

\renewcommand{\filebasisRHS}{ICDM(ur)}
\renewcommand{\filebasisLHS}{Fb.1684(ur)}

\hspace{1.3cm}
\begin{minipage}[t]{\mypwidth}
\begin{tikzpicture}[trim axis left, trim axis right]
\bplotR{\filebasisLHS}{1.0}{3}{\minmaxULHS xlabel={$\Mpt$}, xticklabel pos=right, yticklabels={\algQTree{(e)}, \algQTree{(s)}, \algQPeel{(m)}, \algQPeel{(s)}, \algQPeel{(r)}}, yticklabel pos=left, xticklabels={}}{\resultdirLHS}
\end{tikzpicture}
\end{minipage}
\begin{minipage}[t]{\mypwidth}
\begin{tikzpicture}[trim axis left, trim axis right]
\draw (1,2.35) node[above, font=\scriptsize]{(e)\;\DBLPcf{} \texttt{ICDM} network};
\draw (1,2.05) node[above, font=\scriptsize]{Single-attribute latent profiles sampled at random};
\draw (1,1.8) node[above, font=\scriptsize]{from the uniform distribution};
\bplotR{\filebasisLHS}{1.0}{1}{\minmaxELHS xlabel={$\Mpe$}, xticklabel pos=right, yticklabels={}, xticklabels={}}{\resultdirLHS}
\end{tikzpicture}
\end{minipage}
\begin{minipage}[t]{\mypwidth}
\begin{tikzpicture}[trim axis left, trim axis right]
\bplotR{\filebasisLHS}{1.0}{0}{\minmaxTLHS xlabel={$\Mpc$}, xticklabel pos=right, yticklabel pos=right, yticklabels={}, ylabel={}, xticklabels={}}{\resultdirLHS}
\end{tikzpicture}
\end{minipage}
\hspace{0.55cm}
\begin{minipage}[t]{\mypwidth}
\begin{tikzpicture}[trim axis left, trim axis right]
\bplotR{\filebasisRHS}{1.0}{3}{\minmaxURHS xlabel={$\Mpt$}, xticklabel pos=right, yticklabels={}, yticklabel pos=left, xticklabels={}}{\resultdirRHS}
\end{tikzpicture}
\end{minipage}
\begin{minipage}[t]{\mypwidth}
\begin{tikzpicture}[trim axis left, trim axis right]
\draw (1,2.35) node[above, font=\scriptsize]{(f)\;\texttt{Fb.1684} network};
\draw (1,2.05) node[above, font=\scriptsize]{Single-attribute latent profiles sampled at random};
\draw (1,1.8) node[above, font=\scriptsize]{from the uniform distribution};
\bplotR{\filebasisRHS}{1.0}{1}{\minmaxERHS xlabel={$\Mpe$}, xticklabel pos=right, yticklabels={}, xticklabels={}}{\resultdirRHS}
\end{tikzpicture}
\end{minipage}
\begin{minipage}[t]{\mypwidth}
\begin{tikzpicture}[trim axis left, trim axis right]
\bplotR{\filebasisRHS}{1.0}{0}{\minmaxTRHS xlabel={$\Mpc$}, xticklabel pos=right, yticklabel pos=right, yticklabels={}, ylabel={$D1$}, xticklabels={}}{\resultdirRHS}
\end{tikzpicture}
\end{minipage}

\vspace{0.2em}

\hspace{1.3cm}
\begin{minipage}[t]{\mypwidth}
\begin{tikzpicture}[trim axis left, trim axis right]
\bplotR{\filebasisLHS}{2.0}{3}{\minmaxULHS xticklabel pos=left, yticklabels={\algQTree{(e)}, \algQTree{(s)}, \algQPeel{(m)}, \algQPeel{(s)}, \algQPeel{(r)}}, yticklabel pos=left, xticklabels={}}{\resultdirLHS}
\end{tikzpicture}
\end{minipage}
\begin{minipage}[t]{\mypwidth}
\begin{tikzpicture}[trim axis left, trim axis right]
\bplotR{\filebasisLHS}{2.0}{1}{\minmaxELHS xticklabel pos=left, yticklabels={}, xticklabels={}}{\resultdirLHS}
\end{tikzpicture}
\end{minipage}
\begin{minipage}[t]{\mypwidth}
\begin{tikzpicture}[trim axis left, trim axis right]
\bplotR{\filebasisLHS}{2.0}{0}{\minmaxTLHS xticklabel pos=left, yticklabel pos=right, yticklabels={}, xticklabels={}, ylabel={}}{\resultdirLHS}
\end{tikzpicture}
\end{minipage}
\hspace{0.55cm}
\begin{minipage}[t]{\mypwidth}
\begin{tikzpicture}[trim axis left, trim axis right]
\bplotR{\filebasisRHS}{2.0}{3}{\minmaxURHS xticklabel pos=left, yticklabels={}, yticklabel pos=left, xticklabels={}}{\resultdirRHS}
\end{tikzpicture}
\end{minipage}
\begin{minipage}[t]{\mypwidth}
\begin{tikzpicture}[trim axis left, trim axis right]
\bplotR{\filebasisRHS}{2.0}{1}{\minmaxERHS xticklabel pos=left, yticklabels={}, xticklabels={}}{\resultdirRHS}
\end{tikzpicture}
\end{minipage}
\begin{minipage}[t]{\mypwidth}
\begin{tikzpicture}[trim axis left, trim axis right]
\bplotR{\filebasisRHS}{2.0}{0}{\minmaxTRHS xticklabel pos=left, yticklabel pos=right, yticklabels={}, xticklabels={}, ylabel={$D2$}}{\resultdirRHS}
\end{tikzpicture}
\end{minipage}

\vspace{-0.4em}

\hspace{1.3cm}
\begin{minipage}[t]{\mypwidth}
\begin{tikzpicture}[trim axis left, trim axis right]
\bplotR{\filebasisLHS}{3.0}{3}{\minmaxULHS xticklabel pos=left, yticklabels={\algQTree{(e)}, \algQTree{(s)}, \algQPeel{(m)}, \algQPeel{(s)}, \algQPeel{(r)}}, yticklabel pos=left}{\resultdirLHS}
\end{tikzpicture}
\end{minipage}
\begin{minipage}[t]{\mypwidth}
\begin{tikzpicture}[trim axis left, trim axis right]
\bplotR{\filebasisLHS}{3.0}{1}{\minmaxELHS xticklabel pos=left, yticklabels={}}{\resultdirLHS}
\end{tikzpicture}
\end{minipage}
\begin{minipage}[t]{\mypwidth}
\begin{tikzpicture}[trim axis left, trim axis right]
\bplotR{\filebasisLHS}{3.0}{0}{\minmaxTLHS xticklabel pos=left, yticklabel pos=right, yticklabels={}, ylabel={}}{\resultdirLHS}
\end{tikzpicture}
\end{minipage}
\hspace{0.55cm}
\begin{minipage}[t]{\mypwidth}
\begin{tikzpicture}[trim axis left, trim axis right]
\bplotR{\filebasisRHS}{3.0}{3}{\minmaxURHS xticklabel pos=left, yticklabels={}, yticklabel pos=left}{\resultdirRHS}
\end{tikzpicture}
\end{minipage}
\begin{minipage}[t]{\mypwidth}
\begin{tikzpicture}[trim axis left, trim axis right]
\bplotR{\filebasisRHS}{3.0}{1}{\minmaxERHS xticklabel pos=left, yticklabels={}}{\resultdirRHS}
\end{tikzpicture}
\end{minipage}
\begin{minipage}[t]{\mypwidth}
\begin{tikzpicture}[trim axis left, trim axis right]
\bplotR{\filebasisRHS}{3.0}{0}{\minmaxTRHS xticklabel pos=left, yticklabel pos=right, yticklabels={}, ylabel={$D3$}}{\resultdirRHS}
\end{tikzpicture}
\end{minipage}
\caption{Results for the \probQTeam{} problem on the \DBLPcf{} \texttt{ICDM} and \texttt{Fb.1684} networks with single-attribute latent profiles.}
\label{fig:hists_opiQ_six}
\end{figure}

Looking at random profiles
(Figures~\ref{fig:hists_opiQ_six}c, \ref{fig:hists_opiQ_six}d and \ref{fig:hists_opiQ_six}e) versus eigenvector profiles
(Figures~\ref{fig:hists_opiQ_six}a and \ref{fig:hists_opiQ_six}b), the gap between the
profile-oblivious variants (i.e.\ \algQTree{(e)} and \algQPeel{(r)})
and the profile-aware variants increases. Indeed while the
profile-aware variants generally tend to pick edges with lower tension
at the cost of involving more edges, as discussed earlier, this
tendency is more pronounced when handling eigenvector profiles as
compared to the random distributions. This shows that the
profile-aware variants are clearly suited to exploit the structure
present in eigenvector profiles.

On the other hand, while variations can be observed between the
different random distributions, these differences appear to be limited
when contrasted with the gap that exists between random and eigenvector
profiles. This indicates that the distribution of profile
values has a limited impact compared to the presence of structure.

\parahead{Impact of the network structure}
The profile-aware variants typically achieve the best
performance. Yet, variations in their behavior can be
observed depending for instance on the network structure. 

This is clearly evidenced by Figure~\ref{fig:hists_opiQ_six}f, showing results on the \texttt{Fb.1684} network with single-attribute profiles sampled at random from the uniform distribution.
\texttt{Fb.1684} is an ego-net representing friend lists from Facebook,\!\footnote{\url{http://snap.stanford.edu}} and has a high density, $\delta = 18.07$.
In such high density network, favoring low tension paths as done
by \algQTree{(s)} can result in many more edges in the induced subgraph,
yielding a significantly higher social tension and actually hurting the performance.

\begin{table}
\centering
\caption{Average running times (in seconds) of the algorithms on the \DBLPcf{} \texttt{ICDM} network with latent profiles derived from conferences for solving the \probQTeam{} problem and the \probSTeam{} problem using keywords as skills ($\pm$ standard deviation).}
\label{tab:runtimesQP}
\footnotesize
\begin{tabular}{@{\hspace*{.5em}}c@{\hspace*{0em}}r@{\hspace*{0.2em}}r@{\hspace*{0em}}c@{\hspace*{1.4em}}c@{\hspace*{0em}}r@{\hspace*{0.2em}}r@{\hspace*{0em}}c@{\hspace*{1.4em}}c@{\hspace*{0em}}r@{\hspace*{0.2em}}r@{\hspace*{0em}}c@{\hspace*{1.4em}}c@{\hspace*{0em}}r@{\hspace*{0.2em}}r@{\hspace*{0em}}c@{\hspace*{1.4em}}c@{\hspace*{0em}}r@{\hspace*{0.2em}}r@{\hspace*{0em}}c@{\hspace*{.5em}}}
\toprule
\multicolumn{4}{c}{\algQTree{(e)}} & \multicolumn{4}{c}{\algQTree{(s)}} & \multicolumn{4}{c}{\algQPeel{(s)}} & \multicolumn{4}{c}{\algQPeel{(m)}} & \multicolumn{4}{c}{\algQPeel{(r)}} \\ 
\midrule 
& \\  [-0.75em]
\multicolumn{20}{c}{ \textsc{Solving the \probQTeam{} problem} } \\
 & \multicolumn{18}{c}{\textit{Single-attribute latent profiles}} & \\
 & $0.2$ & $(\pm 0.0)$ &  &  & $3.5$ & $(\pm 0.5)$ &  &  & $147$ & $(\pm 26.6)$ &  &  & $107$ & $(\pm 24.9)$ &  &  & $37.0$ & $(\pm 12.1)$ & \\ [0.6em]
 & \multicolumn{18}{c}{\textit{Multi-attribute latent profiles}} & \\
& $0.2$ & $(\pm 0.0)$ &  &  & $3.3$ & $(\pm 0.3)$ &  &  & $164$ & $(\pm 28.1)$ &  &  & $134$ & $(\pm 29.7)$ &  &  & $38.9$ & $(\pm 11.2)$ & \\
& \\  [-0.75em]
 \cmidrule(lr){4-16}
& \\  [-0.75em]
\multicolumn{20}{c}{ \textsc{Solving the \probSTeam{} problem} } \\
  & \multicolumn{18}{c}{\textit{Single-attribute latent profiles}} & \\
  & $0.7$ & $(\pm 0.3)$ &  &  & $2.8$ & $(\pm 1.2)$ &  &  & $337$ & $(\pm 236)$ &  &  & $261$ & $(\pm 204)$ &  &  & $91.5$ & $(\pm 64.5)$ &  \\ [0.6em]
 & \multicolumn{18}{c}{\textit{Single-attribute latent profiles, $\abs{P}=7$}} & \\
 & $0.9$ & $(\pm 0.1)$ &  &  & $3.1$ & $(\pm 0.9)$ &  &  & $423$ & $(\pm 156)$ &  &  & $351$ & $(\pm 120)$ &  &  & $108$ & $(\pm 40.9)$ & \\  [0.6em]
 & \multicolumn{18}{c}{\textit{Multi-attribute latent profiles}} & \\
 & $0.1$ & $(\pm 0.0)$ &  &  & $2.9$ & $(\pm 1.1)$ &  &  & $127.0$ & $(\pm 46.3)$ &  &  & $108.5$ & $(\pm 35.6)$ &  &  & $33.6$ & $(\pm 13.2)$ & \\ [.6em]
 & \multicolumn{18}{c}{\textit{Multi-attribute latent profiles, $\abs{P}=7$}} & \\
 & $0.1$ & $(\pm 0.0)$ &  &  & $3.3$ & $(\pm 0.6)$ &  &  & $140.8$ & $(\pm 32.7)$ &  &  & $116.9$ & $(\pm 26.9)$ &  &  & $35.5$ & $(\pm 11.7)$ & \\  
\bottomrule
\end{tabular}

\end{table}




\renewcommand{\minmaxCT}{xmin=0.0, xmax=4,}
\renewcommand{\minmaxCE}{xmin=0.0, xmax=16,}
\renewcommand{\minmaxCU}{xmin=0.0, xmax=5,}

\renewcommand{\resultdir}{./xps} 

\begin{figure}[t]
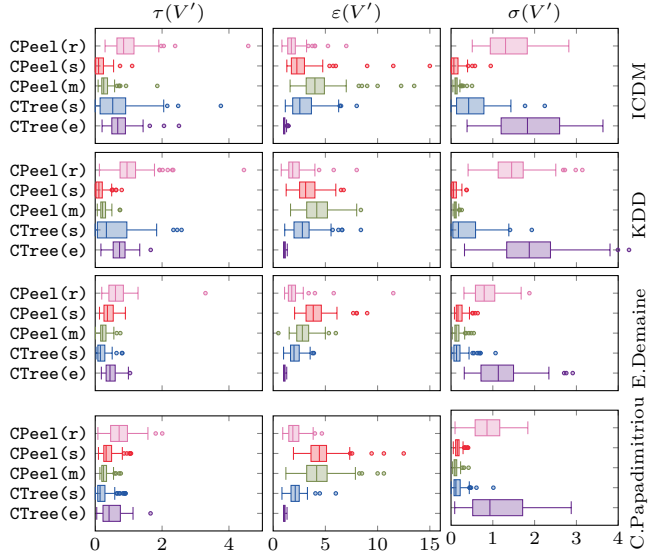

\centering
\bplotBlockC{DBLP.ICDM.(c)}{DBLP.KDD.(c)}{E.Demaine.(c)}{C.Papadimitriou.(c)}{ICDM}{KDD}{E.Demaine}{C.Papadimitriou}
\caption{Results for the \probSTeam{} problem on co-authorship networks from the \DBLPcf{} and \DBLPed{} collections with single-attribute latent profiles derived from keywords using conferences as skills.}
\label{fig:hists_opi_skills}
\end{figure}

\parahead{Running times} 
Indicative running times of the different algorithms for the \probQTeam{} problem on the \DBLPcf{} \texttt{ICDM} network with single-attributes and multi-attributes latent profiles are shown in Table~\ref{tab:runtimesQP}~(top). 
We observe that, also as expected, going from single-attribute to
multi-attribute profiles hardly has any impact on the running times of
our algorithms.

Indicative running times of the algorithms on networks of
varying sizes and densities, with multi-attribute profiles are listed
in Table~\ref{tab:runtimesL}.  As expected, the tree-based algorithms
are significantly faster than the top-down algorithms, up to two
orders of magnitude, and scale much better.  

\begin{table*}
\centering
\caption{Average running times (in seconds) of the algorithms ($\pm$ standard deviation) solving the \probQTeam{} problem on networks of varying number of vertices ($\left|V\right|$), edges ($\left|E\right|$) and average degree densities ($\density$).}
\label{tab:runtimesL}
\footnotesize
\begin{tabular}{@{\hspace*{0.2em}}l@{\hspace*{3em}}r@{\hspace*{0.6em}}r@{\hspace*{0.6em}}r@{\hspace*{2.5em}}c@{\hspace*{0em}}r@{\hspace*{0.2em}}r@{\hspace*{0em}}c@{\hspace*{0.65em}}c@{\hspace*{0em}}r@{\hspace*{0.2em}}r@{\hspace*{0em}}c@{\hspace*{0.75em}}c@{\hspace*{0em}}r@{\hspace*{0.2em}}r@{\hspace*{0em}}c@{\hspace*{0.65em}}c@{\hspace*{0em}}r@{\hspace*{0.2em}}r@{\hspace*{0em}}c@{\hspace*{0.65em}}c@{\hspace*{0em}}r@{\hspace*{0.2em}}r@{\hspace*{0em}}c@{\hspace*{0.2em}}}
\toprule
Network & $\abs{V}$ & $\abs{E}$ & $\delta$ & \multicolumn{4}{c}{\algQTree{(e)}} & \multicolumn{4}{c}{\algQTree{(s)}} & \multicolumn{4}{c}{\algQPeel{(s)}} & \multicolumn{4}{c}{\algQPeel{(m)}} & \multicolumn{4}{c}{\algQPeel{(r)}} \\
\midrule 
\IMDB{} \texttt{Kassovitz} & $96$ & $262$  & $2.73$ &  & $0.0$ & $(\pm 0.0)$ &  &  & $0.0$ & $(\pm 0.0)$ &  &  & $0.1$ & $(\pm 0.0)$ &  &  & $0.1$ & $(\pm 0.0)$ &  &  & $0.0$ & $(\pm 0.0)$ &  \\
\IMDB{} \texttt{J.Cameron} & $278$ & $8\,98$ & $3.23$ &  & $0.0$ & $(\pm 0.0)$ &  &  & $0.1$ & $(\pm 0.0)$ &  &  & $0.5$ & $(\pm 0.1)$ &  &  & $0.4$ & $(\pm 0.1)$ &  &  & $0.1$ & $(\pm 0.0)$ &  \\
\IMDB{} \texttt{WarnerBros 1970s} & $225$ & $1\,599$ & $7.11$ &  & $0.0$ & $(\pm 0.0)$ &  &  & $0.1$ & $(\pm 0.0)$ &  &  & $0.9$ & $(\pm 0.1)$ &  &  & $0.7$ & $(\pm 0.1)$ &  &  & $0.1$ & $(\pm 0.1)$ &  \\
\IMDB{} \texttt{Forman} & $395$ & $1\,638$ & $4.15$ &  & $0.0$ & $(\pm 0.0)$ &  &  & $0.2$ & $(\pm 0.0)$ &  &  & $1.3$ & $(\pm 0.2)$ &  &  & $1.0$ & $(\pm 0.2)$ &  &  & $0.2$ & $(\pm 0.1)$ &  \\
\IMDB{} \texttt{Eastwood 2000s} & $449$ & $2\,307$ & $5.14$ &  & $0.0$ & $(\pm 0.0)$ &  &  & $0.2$ & $(\pm 0.0)$ &  &  & $2.2$ & $(\pm 0.4)$ &  &  & $1.8$ & $(\pm 0.4)$ &  &  & $0.4$ & $(\pm 0.2)$ &  \\
\IMDB{} \texttt{F.F.Coppola} & $678$ & $6\,306$ & $9.30$ &  & $0.0$ & $(\pm 0.0)$ &  &  & $0.7$ & $(\pm 0.1)$ &  &  & $8.8$ & $(\pm 1.6)$ &  &  & $6.4$ & $(\pm 1.2)$ &  &  & $1.4$ & $(\pm 0.6)$ &  \\
\IMDB{} \texttt{WarnerBros 2000s} & $1\,032$ & $7\,279$ & $7.05$ &  & $0.1$ & $(\pm 0.0)$ &  &  & $0.7$ & $(\pm 0.0)$ &  &  & $14.4$ & $(\pm 2.5)$ &  &  & $10.1$ & $(\pm 2.2)$ &  &  & $2.4$ & $(\pm 1.1)$ &  \\
\DBLPed{} \texttt{E.Demaine} &$2\,234$ & $7\,701$ & $3.45$ &  & $0.1$ & $(\pm 0.0)$ &  &  & $2.8$ & $(\pm 0.3)$ &  &  & $75.7$ & $(\pm 12.3)$ &  &  & $60.9$ & $(\pm 13.0)$ &  &  & $19.2$ & $(\pm 6.2)$ &  \\
\IMDB{} \texttt{Paramount 2000s} & $1\,097$ & $8\,469$ & $7.72$ &  & $0.1$ & $(\pm 0.0)$ &  &  & $0.8$ & $(\pm 0.1)$ &  &  & $17.8$ & $(\pm 3.0)$ &  &  & $12.1$ & $(\pm 2.5)$ &  &  & $3.2$ & $(\pm 1.2)$ &  \\
\DBLPed{} \texttt{C.Papadimitriou} & $2\,613$ & $9\,472$ & $3.62$ &  & $0.1$ & $(\pm 0.0)$ &  &  & $3.2$ & $(\pm 0.3)$ &  &  & $114.6$ & $(\pm 20.0)$ &  &  & $91.6$ & $(\pm 22.0)$ &  &  & $31.6$ & $(\pm 9.6)$ &  \\
\DBLPcf{} \texttt{ICDM} & $2\,795$ & $10\,280$ & $3.68$ &  & $0.2$ & $(\pm 0.0)$ &  &  & $3.3$ & $(\pm 0.3)$ &  &  & $163.9$ & $(\pm 28.1)$ &  &  & $133.9$ & $(\pm 29.7)$ &  &  & $38.9$ & $(\pm 11.2)$ &  \\
\DBLPcf{} \texttt{KDD} & $2\,737$ & $11\,072$ & $4.05$ &  & $0.2$ & $(\pm 0.0)$ &  &  & $3.5$ & $(\pm 0.2)$ &  &  & $166.8$ & $(\pm 27.8)$ &  &  & $136.7$ & $(\pm 28.1)$ &  &  & $36.0$ & $(\pm 13.0)$ &  \\
\DBLPed{} \texttt{P.Yu} & $4\,596$ & $13\,250$ & $2.88$ &  & $0.2$ & $(\pm 0.0)$ &  &  & $4.4$ & $(\pm 0.3)$ &  &  & $291.3$ & $(\pm 56.3)$ &  &  & $242.3$ & $(\pm 43.1)$ &  &  & $68.6$ & $(\pm 20.5)$ &  \\
\IMDB{} \texttt{Paramount} & $1\,952$ & $28\,992$ & $14.85$ &  & $0.3$ & $(\pm 0.0)$ &  &  & $2.6$ & $(\pm 0.1)$ &  &  & $116.2$ & $(\pm 17.3)$ &  &  & $49.3$ & $(\pm 11.1)$ &  &  & $17.7$ & $(\pm 7.8)$ &  \\
\IMDB{} \texttt{WarnerBros} & $2\,111$ & $32\,166$ & $15.24$ &  & $0.3$ & $(\pm 0.1)$ &  &  & $3.0$ & $(\pm 0.2)$ &  &  & $139.1$ & $(\pm 19.7)$ &  &  & $57.2$ & $(\pm 13.0)$ &  &  & $22.8$ & $(\pm 9.1)$ &  \\
\IMDB{} \texttt{WB+Paramount+Fox} & $5\,758$ & $178\,741$ & $31.04$ &  & $1.4$ & $(\pm 0.2)$ &  &  & $15.4$ & $(\pm 1.0)$ &  &  & $2192.3$ & $(\pm 346.6)$ &  &  & $670.5$ & $(\pm 168.1)$ &  &  & $281.6$ & $(\pm 101.6)$ &  \\
\bottomrule

\end{tabular}
\end{table*}

\subsection{Team formation with chosen skills}
\label{sec:exp_skills}
Next, we turn to the \probSTeam{} problem, the problem variant where we are
given a project requiring a set of skills and asked to find a team which is
connected, has low social tension and covers the chosen skills.

\parahead{Generating sets of skills}
For the experiments with the {\probSTeam} problem we consider the co-authorship networks from
the \DBLPcf{} and \DBLPed{} collections.  Here, we take the conferences to be the skills associated with
individuals.
Specifically, each conference represents a skill which a researcher is
considered to possess if he has published at least four papers in that
particular conference. A project then consists of a subset of conferences.
Thus, forming a team to fulfill the project can be thought of as finding
a group of researchers that span the sub-areas of computer science represented by these conferences.


For each network in \DBLPcf{} and \DBLPed{}, we randomly sampled $80$ projects
(subsets of 3 to 13 conferences present in that network). Following the two-step
procedure described in Section~\ref{sec:variants}, we then used our five
algorithm variants to solve the \probSTeam{} problem for each dataset (a
network and its associated keyword-based latent profiles) and each project.
Results are presented in Figure~\ref{fig:hists_opi_skills}.

\parahead{Characteristics of the reported teams}
We observe the same general trend as for \probQTeam{}
 (Section~\ref{sec:exp_seeds}). That is, profile-aware variants
\algQPeel{(s)}, \algQPeel{(m)}, and \algQTree{(s)} outperform the
other variants, although \algQPeel{(r)} and \algQTree{(e)} report
subgraphs with fewer edges. In the \DBLPcf{} networks, however, we see
an increased variability in the behavior of the \algQTree{(s)} variant
and a notable performance of \algQPeel{(m)}. This latter algorithm now appears to return solutions
that have both small size and low edge weights.

\parahead{Running times} 
Indicative running times of the different algorithms for this problem are shown in Table~\ref{tab:runtimesQP}~(bottom). 
To allow comparison against the running times for \probQTeam{} reported in the top half of the table, we include times restricted to runs with projects requiring seven skills, i.e.\ $\abs{P}=7$. Note, that while this implies that the number of seed nodes in the first run equals seven, the number of seed nodes in the second run might vary. 

While the running times for the two-step procedure are expectedly larger than those of the basic algorithms, the variance is also much greater. In particular, this is because while two runs are often necessary to find a connected solution, it can happen that a single run suffices but such cases cannot be easily identified at the outset.


\section{Conclusions}\label{sec:conclusions}
Problems related to community search and team formation have
multiple applications in online social media and collaboration networks.
In this paper, we add a new modeling angle to these two classes of problems.
The key characteristic of our model is that each node of the social network is not only characterized by its connections
and its skills, but also by its profile. These profiles, which change dynamically through a conformation process,
give rise to social tension in the network.  Given this model, we define the 
{\probQTeam} and {\probSTeam} problems, where the goal is to identify
a set of connected individuals that define a low-tension subgraph. 
Such problems arise both in social network and social media mining as well as
in human-resource management, where the goal is to find a set of workers who are not only connected, but also  will have a potentially  fluid collaboration.
The  contributions of our paper include the formal definition of these problems and the design of algorithms
for solving them effectively in practice.  Our experimental results with real data from social and collaboration
networks highlight the characteristic behavior of the different algorithms variants and illustrate the effect of network structure and profile distribution on the algorithms' relative performance.
Finally, our work enables  future research 
combining subgraph mining with dynamic processes occurring
among the nodes.  


\smallskip
\noindent
{\bf Acknowledgements.}
Most of the work was done while Esther Galbrun and Behzad Golshan were at Boston University.
Aristides Gionis is supported by 
the Finnish Funding Agency for Innovation TEKES (project ``Re:Know''), 
the Academy of Finland (project ``Nestor''), and
the EU H2020 Program (project ``SoBigData''). This research was funded
by NSF grants: IIS 1320542, IIS 1421759 and CAREER 1253393 as well as a gift
from Microsoft.

\bibliographystyle{abbrv}
\bibliography{opinions} 

\end{document}